\documentclass{aastex63}

\received{???}
\revised{???}
\accepted{???}
\submitjournal{ApJ}

\shorttitle{Suprathermal Protons}
\shortauthors{Young et al.}
\watermark{Draft}
\graphicspath{{./}{figures/}}

\begin{document}

\title{Suprathermal Proton Spectra at Interplanetary Shocks in 3-D Hybrid Simulations}

\correspondingauthor{Matthew A. Young}
\email{myoung.space.science@gmail.com}

\author[0000-0003-2124-7814]{Matthew A. Young}
\affiliation{University of New Hampshire \\
Morse Hall \\
8 College Road \\
Durham, NH, 03824, USA}

\author{Bernard J. Vasquez}
\affiliation{University of New Hampshire \\
Morse Hall \\
8 College Road \\
Durham, NH, 03824, USA}

\author{Harald Kucharek}
\affiliation{University of New Hampshire \\
Morse Hall \\
8 College Road \\
Durham, NH, 03824, USA}

\author[0000-0002-1890-6156]{No\'e Lugaz}
\affiliation{University of New Hampshire \\
Morse Hall \\
8 College Road \\
Durham, NH, 03824, USA}



\begin{abstract} 

Interplanetary shocks are one of the proposed sources of suprathermal ion populations (i.e.,  ions with energies of a few times the solar wind energy). Here, we present results from a series of three-dimensional hybrid simulations of collisionless shocks in the solar wind. We focus on the influence of the shock-normal angle, $\theta_{Bn}$, and the shock speed, $V_s$, on producing protons with energies a few to hundreds of times the thermal energy of the upstream plasma. The combined effects of $\theta_{Bn}$ and $V_s$ result in shocks with Alfv{\'e}n Mach numbers in the range 3.0 to 6.0 and fast magnetosonic Mach numbers in the range 2.5 to 5.0, representing moderate to strong interplanetary shocks. We find that $\theta_{Bn}$ largely organizes the shape of proton energy spectra while shock speed controls acceleration efficiency. All shocks accelerate protons at the shock front but the spectral evolution depends on $\theta_{Bn}$. Shocks with $\theta_{Bn} \geq 60^\circ$ produce isolated bursts of suprathermal protons at the shock front while shocks with $\theta_{Bn} \leq 45^\circ$ create suprathermal beams upstream of the shock. Downstream proton energy spectra have exponential or smoothed broken power-law forms when $\theta_{Bn} \geq 45^\circ$, and a single power-law form when $\theta_{Bn} \leq 30^\circ$. Protons downstream of the strongest shocks have energies at least 100 times the upstream thermal energy, with $\theta_{Bn} \leq 30^\circ$ shocks producing the highest energy protons and $\theta_{Bn} \geq 60^\circ$ shocks producing the largest number of protons with energies at least a few times the thermal energy.

\end{abstract}





\section{Introduction} 
\label{sec:intro}

The presence of a suprathermal ion population in the solar wind has been known for decades \citep{Gosling_etal-1981-InterplanetaryIonsDuring}, but its physical origin remains an open question. By suprathermal ions, we mean ions with speeds of 1.4-10 times the solar wind speed or energies in the 2-100 keV per nucleon (keV nuc.${}^{-1}$) range. The relevance of suprathermal ions to the field of heliophysics is both academic and operational. In the former case, it is an intriguing topic within the scope of collisionless plasma theory \citep[e.g.][]{Pierrard_Lazar-2010-KappaDistributionsTheory}. In the latter case, it is of great importance to predicting the arrival and flux of energetic particles at locations where they can threaten astronauts or spacecraft \citep[e.g.][]{Laming_etal-2013-RemoteDetectionSuprathermal}. Suprathermal ions provide the seed population for particles that get accelerated to MeV-range energies, and it follows that the spectrum of that seed population should affect the resultant spectrum of energetic particles. This paper provides insight into the role that shocks play in producing suprathermal protons from the bulk solar-wind population. It investigates proton spectra upstream of, at, and downstream of a variety of shocks, with attention to their role in producing a seed population for energetic particle events.

\subsection{Motivation}
\label{sec:motivation}

There is strong evidence that interplanetary (IP) shocks create solar energetic particles (SEPs) and energetic storm particles (ESP) with energies of a few MeV nuc.${}^{-1}$ and higher. Most observations of shock-driven SEPs are associated with coronal mass ejections (CMEs) (e.g, \citet{Reames-1993-NonThermalParticles}, \citet{Cane-1995-StructureEvolutionInterplanetary}, and \citet{Zank_etal-2000-ParticleAccelerationCoronal}), but stream interaction regions (SIRs) and corotating interaction regions (CIRs) can also accelerate particles to MeV energies (e.g., \citet{McDonald_etal-1976-InterplanetaryAccelerationEnergetic}, \citet{Barnes_Simpson-1976-EvidenceInterplanetaryAcceleration}, and \citet{Richardson_etal-1993-CorotatingMeVIon}). \citet{Giacalone-2012-EnergeticChargedParticles} analyzed the 18 strong shocks present in \textit{ACE} observations from 1998 to 2003 and found that the intensity of 47-65 keV protons associated with each shock was enhanced over the previous day's level. \citet{Kahler_Ling-2019-SuprathermalIonBackgrounds} examined signatures of suprathermal H and He at 100 keV nuc.${}^{-1}$ and 1 MeV nuc.${}^{-1}$ prior to SEP events with 10 MeV nuc.${}^{-1}$ and found a longitudinal correlation that led them to propose an acceleration mechanism involving quasi-perpendicular shocks. Observations of heavier elements in shock-driven SEPs, such as Fe at 10-40 MeV nuc.${}^{-1}$, also exhibit a dependence on the intensity of an ambient lower-energy seed population \citep{Mewaldt_etal-2012-DependenceSolarEnergetic}. Beyond the simple CME-driven shock picture, CME-CME interactions may provide a multi-step acceleration process in which the first CME creates a seed population which the second CME can accelerate to tens of MeV \citep{Schmidt_Cargill-2004-NumericalStudyTwo, Li_Zank-2005-MultipleCMEsLarge, Li_etal-2012-TwinCMEScenario, Lugaz_etal-2017}. It is often difficult or impossible to know the state of the coronal/solar-wind plasma through which a transient shock or compression passes, but the presence of a previously enhanced particle population appears to be a determining factor in whether a CME shock can produce high-intensity SEP events \citep{Kahler-2001-CorrelationBetweenSolar}.

Many works propose diffusive shock acceleration (DSA) \citep{Axford_etal-1977-AccelerationCosmicRays, Blandford_Ostriker-1978-ParticleAccelerationAstrophysical, Bell-1978-AccelerationCosmicRays1,Bell-1978-AccelerationCosmicRays2, Drury-1983-IntroductionTheoryDiffusive, Scholer-1985-DiffusiveAcceleration,Kucharek_Scholer-1991-OriginDiffuseSuprathermal} as a mechanism for producing SEPs. Particles involved in DSA gain energy through a first-order Fermi-type process \citep{Fermi-1949-OriginCosmicRadiation} involving repeated shock crossings due to reflections from turbulent magnetic structures. It is an attractive explanation for particle acceleration throughout the universe because it naturally creates power-law distributions, which are ubiquitous in observations. However, the DSA process requires that a particle have a minimum energy in order to participate, since it must have sufficient momentum to interact with a downstream structure and return upstream. The issue of how particles acquire this minimum energy (or momentum) is known as the \emph{injection} problem. Assuming DSA is responsible for accelerating a suprathermal seed population to energies greater than 10 MeV near the Sun, the \citet{Kahler_Ling-2019-SuprathermalIonBackgrounds} results suggest that the threshold energy injection is at most 100 keV. 

Shock-drift acceleration (SDA) \citep{Jokipii-1982-ParticleDriftDiffusion, Armstrong_etal-1985-ShockDriftAcceleration, Anagnostopoulos-1994-DominantAccelerationProcesses} is a candidate mechanism for injecting ions into first-order Fermi acceleration. In SDA, the sharp magnetic-field gradient across a shock front causes particles to drift across the shock front, where the electric field induced by the shock motion can accelerate them. The initial energy of a pre-existing particle distribution likely affects the accelerated distribution, since a given particle spends a finite amount of time in the acceleration region, but it does not formally require a minimum initial energy. Therefore, SDA should be capable of directly accelerating particles from the thermal distribution.

In a survey of 74 out of 400 IP shocks driven by CMEs observed during solar cycle 23 by \textit{ACE} and \textit{Wind}, \citet{Desai_etal-2012-IonAccelerationNear} found that 87\% of the shocks exhibited a mix of first-order Fermi acceleration and shock-drift acceleration, rather than a clear signature of one or the other. Of the remaining ten shocks, four with $\theta_{Bn} > 70^\circ$ exhibited signatures of shock-drift acceleration and six with $\theta_{Bn} < 70^\circ$ exhibited signatures of first-order Fermi acceleration. The significant association between coherent magnetic structures and the flux of 0.047-4.75 MeV particles in \textit{ACE} data further illustrates the complexity of the particle acceleration process \citep{Tessein_etal-2013-AssociationSuprathermalParticles,Tessein_etal-2015-EffectCoherentStructures}. In fact, there is a variety of physical mechanisms that might generate a population of seed particles which a shock could subsequently accelerate \citep{Laming_etal-2013-RemoteDetectionSuprathermal}. This work demonstrates the ability of an initial shock with a Mach number typical of IP shocks in the inner heliosphere to accelerate thermal particles to suprathermal energies, thereby providing the seed population for a subsequent energization process.

\subsection{Previous simulations}
\label{sec:Previous simulations}

Previous numerical simulations have provided insight into the generation of suprathermal ion populations, especially as the supporting technology has advanced. One common simulation technique is the hybrid technique, which self-consistently models ions as particles (that is, kinetically) and electrons as a fluid. Another common simulation technique is the test-particle technique, which allows existing fields to affect particle dynamics but does not self-consistently allow particles to modify fields. Both methods have their advantages and disadvantages.

A number of researchers \citep{Quest-1988,Scholer_Terasawa-1990,Kucharek_Scholer-1991-OriginDiffuseSuprathermal,Giacalone_etal-1992,Sugiyama_Terasawa-1999-ScatterFreeIon,Su_etal-2012-ParticleAccelerationGeneration} investigated the initial energization at parallel shocks based on the results of self-consistent hybrid simulations. More recent one-dimensional (1-D) simulations of a parallel shock by \citet{Giacalone-2004-LargeScaleHybrid}, performed using four different spatial domains, produced results that were qualitatively consistent with diffusive theories of acceleration in which energetic particles couple to self-generated magnetic fluctuations. Classical DSA operates most efficiently at parallel shocks since that geometry encourages particles to repeatedly cross the shock front, which leads to energy gain. However, two-dimensional (2-D) test-particle simulations suggest that shocks readily accelerate low-energy particles to high energies irrespective of the mean shock-normal angle, $\langle\theta_{Bn}\rangle$ \citep{Giacalone-2005-ParticleAccelerationShocks}. Those simulations showed that the acceleration rate is larger for perpendicular shocks, and subsequent 2-D hybrid simulations showed that perpendicular shocks can directly energize a fraction of the thermal population by reflecting particles back upstream along multiply connected field lines \citep{Giacalone-2005-EfficientAccelerationThermal}. The results from this combination of test-particle and hybrid simulations suggests that perpendicular shocks may bypass the minimum-energy requirement of DSA.

\citet{Caprioli_Spitkovsky-2014-SimulationsIonAcceleration1} modeled primarily 2-D shocks at seven shock-normal angles with a hybrid code and found that acceleration efficiency is much lower for perpendicular shocks than for parallel shocks. This is in contrast to observations by \citet{Reames-2012-ParticleEnergySpectra}, as well as to the results of hybrid simulations by \citet{Giacalone-2005-EfficientAccelerationThermal}, though test-particle simulations with upstream magnetic turbulence by \citet{Giacalone-2005-ParticleAccelerationShocks} displayed a weak dependence on $\theta_{Bn}$ with the same trend as described by \citet{Caprioli_Spitkovsky-2014-SimulationsIonAcceleration1}.

\citet{Caprioli_Spitkovsky-2014-SimulationsIonAcceleration2} and \citet{Caprioli_Spitkovsky-2014-SimulationsIonAcceleration3} each used a subset of the original 2-D simulation runs to analyze the roles of magnetic-field amplification and particle diffusion, respectively. \citet{Caprioli_etal-2015-SimulationsTheoryIon} built on the 2-D results of \citet{Caprioli_Spitkovsky-2014-SimulationsIonAcceleration1,Caprioli_Spitkovsky-2014-SimulationsIonAcceleration2,Caprioli_Spitkovsky-2014-SimulationsIonAcceleration3} by constructing a minimal model of ion injection into DSA that accounts for shock reformation when the shock-normal angle is small. Although most of the results of \citet{Caprioli_Spitkovsky-2014-SimulationsIonAcceleration1,Caprioli_Spitkovsky-2014-SimulationsIonAcceleration2,Caprioli_Spitkovsky-2014-SimulationsIonAcceleration3} focus on 2-D simulation runs, Section 8 of \citet{Caprioli_Spitkovsky-2014-SimulationsIonAcceleration1} describes a set of 3-D simulation runs at three shock-normal angles. However, much of the focus in their work is on Mach numbers that are more appropriate to astrophysical applications such as supernovae than to IP shocks.

Three-dimensional (3-D) simulations are necessary for accurately modeling charged-particle dynamics in the presence of a magnetic field, where cross-field diffusion is important. \citet{Jokipii_etal-1993-PerpendicularTransport} and \citet{Jones_etal-1998-ChargeParticleMotion} analytically showed that suppressing (or ignoring) at least one coordinate of the magnetic field artificially restricts each particle's motion to the field line on which it started. \citet{Giacalone_Jokipii-1994-ChargedParticleMotion} built upon the work of \citet{Jokipii_etal-1993-PerpendicularTransport} to show that the cross-field motion suppressed in 1- and 2-D simulations actually occurs in 3-D simulations. They used a Kolmogorov-like spectrum for the fluctuating magnetic field. In 3-D simulations, particles diffuse both along and across the field; in 2-D simulations, they stay tied to their original field line and only diffuse via field-line mixing. Subsequent simulations of magnetic turbulence driven by an ion/ion beam instability \citep{Kucharek_etal-2000-ThreeDimensionalSimulation} and of a quasi-perpendicular shock \citep{Giacalone_Ellison-2000-ThreeDimensionalNumerical} confirmed the importance to cross-field diffusion of including all three dimensions. One result from 1- and 2-D simulations that appears to persists in 3-D is that the subset of particles in the high-energy tail downstream of the shock gain their initial energy at the shock \citep{Guo_Giacalone-2013-AccelerationThermalProtons}.

\subsection{Previous observations}
\label{sec:Previous observations}

The above simulation results demonstrate that shocks can theoretically accelerate thermal ions to high energies, perhaps even eliminating the injection problem in certain cases, but the relative paucity of 3-D simulations with shock parameters appropriate to the solar wind (e.g., Mach numbers less than 4) makes this result far from conclusive. Furthermore, numerous satellite observations suggest that SEP and ESP events draw from a seed population of suprathermal ions.

Abundances of trace ions and the composition of heavy ions in populations of accelerated particles correlate better with the suprathermal population than the bulk solar wind --- evidence that the former provides the seed population for energetic particle events \citep{Mason_etal-2005-EnergeticParticlesAccelerated}. Furthermore, the suprathermal ion composition at 1~au is dynamic: Ions accelerated in SEP events dominate it during solar maximum conditions while suprathermal solar wind ions and/or those accelerated during CIRs dominate it during solar minimum conditions \citep{Desai_etal-2006-SolarCycleVariations, Dayeh_etal-2017-OriginPropertiesQuiet}. \citet{Desai_etal-2003-EvidenceSuprathermalSeed} and \citet{Desai_etal-2006-SeedPopulationEnergetic} also presented compelling evidence that the source material for CME-driven shocks comes from the suprathermal tail of the solar wind, rather than the bulk plasma.

In a survey of daily averages of suprathermal Fe density and Fe fluence in solar energetic particles (SEPs) from \textit{ACE} data, \citet{Mewaldt_etal-2012-DependenceSolarEnergetic} found that the pre-existing suprathermal population appeared to limit the maximum SEP fluence and that large SEP fluence only occurred when there had been a pre-existing high suprathermal density. \citet{Mason_etal-2008-AbundancesEnergySpectra} found that heavy-ion spectra in CIRs have similar shapes across species, leading to reasonably constant relative abundances over the observed energy range. They took this as evidence that energetic particles in CIRs are accelerated from a suprathermal ion pool that includes remnant suprathermal ions from previous impulsive SEP events, pickup ions, and heated solar-wind ions.

Energy and velocity spectra of particle distributions in the solar wind typically follow power-law forms at suprathermal and higher energies \citep{Mewaldt_etal-2001-LongTermFluences} and \textit{in situ} observations of suprathermal ions yield differential intensities with a wide range of power-law indices. \citet{Mason_etal-2008-AbundancesEnergySpectra} observed power laws in heavy-ion spectra with indices $\sim 2.51$ below 1 MeV/nuc, which rolled over to power laws with indices $\sim 4.47$ above 1 MeV nuc.${}^{-1}$ \citet{Dayeh_etal-2009-CompositionSpectralProperties} and \citet{Desai_etal-2010-OriginQuietTime} reported spectral indices of 1.27 to 2.29 in quiet-time \textit{ACE} observations while the multi-year heavy-ion fluences reported by \citet{Mewaldt_etal-2001-LongTermFluences} appeared to follow $dJ/dE \propto E^{-2}$.

\subsection{Overview of this work}
\label{sec:Overview of this work}

This paper presents proton energy spectra self-consistently produced by shocks generated in 3-D hybrid plasma simulations of the solar wind. The simulation runs range in shock-normal angle, $\theta_{Bn}$, from parallel to perpendicular, at two values of the upstream flow speed, $V_1$. The resultant shock speeds fall into two groups separated by $V_1$, with intra-group variation determined by $\theta_{Bn}$. Considering the full range of $\theta_{Bn}$ is necessary for developing a complete picture of how seed populations contribute to energetic particle events, since heliographic longitude and observer connectivity can drastically affect the temporal evolution of particle intensity over time at a given location \citep{Cane_etal-1988-RoleInterplanetaryShocks}. Theoretical work by \citet{Tylka_Lee-2006-ModelSpectralCompositional}, and subsequent simulations by \citet{Sandroos_Vainio-2007-SimulationResultsHeavy}, showed how variation in the suprathermal seed population and evolution of the shock-normal angle can account for the high variability in SEP spectra and composition. 

Considering two values of $V_1$ helps determine the effect that shock speed has on the downstream proton spectrum. Observations of energetic He ions in a large sample of IP shocks suggest that significant acceleration and high particle intensity depend strongly upon shock speed. They also depend, albeit less strongly, on compression ratio \citet{Reames-2012-ParticleEnergySpectra}. A study by \citet{Berdichevsky_etal-2000-InterplanetaryFastShocks} of shocks in \textit{Wind} data found that fast-forward shocks had a mean magnetosonic Mach number of $1.39 \pm 0.34$, while the more recent comprehensive statistical analysis of over 600 IP shocks presented by \citet{Kilpua_etal-2015-PropertiesDriversFast} found a median magnetosonic Mach number of 2.1 with an inter-quartile range of 1.20 and a maximum around 4. The simulated shocks in this work therefore represent moderate to strong IP shocks with respect to typical values observed at 1 au. In the solar corona, higher CME speeds may result in higher Mach numbers than at 1 au, despite the fact that the coronal magnetosonic speed is also higher. A recent comparison of multiple techniques by \citet{Maguire_etal-2020-EvolutionAlfvenMach} found Alfv{\'e}n Mach numbers of coronal shocks in the range 1.5 to 4, but \citet{Kozarev_etal-2019-EarlyStageSolar} found coronal Alfv{\'e}n Mach numbers as high as 10.

We present results from 3-D simulations of shocks with Mach numbers $\lesssim 6$, which are more applicable to the IP medium than the 3-D simulations presented in \citet{Caprioli_Spitkovsky-2014-SimulationsIonAcceleration1}. This set of simulation runs also has greater precision in $\theta_{Bn}$ than previous 3-D simulations. Furthermore, it uses roughly six times as many macro-particles per cell as \citet{Caprioli_Spitkovsky-2014-SimulationsIonAcceleration1} and twice as many as \citet{Guo_Giacalone-2013-AccelerationThermalProtons}, thereby significantly improving the resolution of dynamical proton quantities such as energy distribution. Section \ref{sec:Numerical Model} describes the numerical model and relevant input parameters, section \ref{sec:Results} presents results from the simulation runs, section \ref{sec:Discussion} discusses those results in light of previous theoretical and observational work, and section \ref{sec:Conclusion} concludes the paper.

\section{Numerical Model}
\label{sec:Numerical Model}

This work used a plasma model currently under development at the University of New Hampshire. It is a hybrid, collisionless, electromagnetic code designed to model shocks and turbulence in the IP medium. The hybrid scheme, which models ions as particles and electrons as a quasi-neutralizing fluid, captures kinetic ion dynamics without needing to resolve the electron plasma frequency or the Debye length, allowing it to take larger spatial and temporal steps than a fully kinetic code. The code combines the numerical model described in \citet{Vasquez-1995-SimulationStudyRole}, \citet{Vasquez_etal-2014-ThreeDimensionalHybrid}, and \citet{Vasquez-2015-HeatingRateScaling} with the numerical model described in \citet{Kucharek_etal-2000-ThreeDimensionalSimulation}. It uses time-advance schemes similar to those described by \citet{Terasawa_etal-1986-DecayInstabilityFinite} and \citet{Matthews-1994-CurrentAdvanceMethod} and is efficiently parallelized with the Message Passing Interface (MPI). The code self-consistently generates a moving shock by injecting a Maxwellian ion beam through the boundary at $x=0$ and allowing it to reflect from the boundary at $x=L_x$ \citep[cf.][]{Winske_Omidi-1996-NonspecialistsGuideKinetic}. The $y$ and $z$ dimensions are periodic. The code employs a finite difference approach to calculating field quantities in order to permit the non-periodic injection/reflection boundary conditions in the $x$ dimension. 

\subsection{Simulation parameters}
\label{sec:Simulation parameters}

The simulation normalizes number density, $n$, and magnetic intensity, $B$, to upstream values. Where necessary, a subscript ``1'' will denote an upstream value (e.g., $B_1$ for upstream magnetic intensity). It thus normalizes fluid and particle velocities to the upstream Alfv\'en speed, $V_{A,1} = B_1/\sqrt{\mu_0m_pn_1}$. The simulation builds distributions of speed and energy for each particle species by binning each particle's (normalized) peculiar speed and corresponding energy as follows: The peculiar \emph{velocity} is defined as $\delta \mathbf{v} \equiv \tilde{\mathbf{v}} - \mathbf{V}$, where $\tilde{\mathbf{v}}$ is the total particle velocity and $\mathbf{V}$ is the bulk (i.e., average) species velocity. The peculiar speed of the $j^{th}$ component is thus $\delta v_j = \tilde{v}_j - V_j$ and the total peculiar speed (or simply the ``peculiar speed'') is $\delta v = \sqrt{\delta v_x^2 + \delta v_y^2 + \delta v_z^2}$. Finally, this work defines the corresponding normalized variables for each species as $\mathbf{v} = \delta \mathbf{v}/V_{A,1}$, $v_j = \delta v_j/V_{A,1}$, and $v = \delta v/V_{A,1}$. Although it is possible to simulate multiple ion species, the runs presented here used protons as the only ion species.


From the definitions of $\delta v_j$ and $\delta v$, we can define the energy associated with the $j^{th}$ component of the peculiar velocity as $\delta \varepsilon_j \equiv m_p\left(\delta v_j\right)^2/2$, and the energy associated with the total peculiar speed (or simply the ``peculiar energy'') as $\delta \varepsilon \equiv m_p\left(\delta v\right)^2/2$. By defining the upstream Alfv\'en energy, $E_{A,1} = m_pV_{A,1}^2/2$, we can express the normalized peculiar energies as $\varepsilon_j = \delta \varepsilon_j/E_{A,1} = \left(\delta v_j/V_{A,1}\right)^2 = v_j^2$ and $\varepsilon = \delta \varepsilon/E_{A,1} = \left(\delta v/V_{A,1}\right)^2 = v^2$. The simulation runs presented here binned $\varepsilon_j$ and $\varepsilon$ from 0 to 1600, with bin width $\Delta \varepsilon = 0.8$. Assuming $V_{A,1} = 5 \times 10^4$ m/s in the solar wind at 1~au, $E_{A,1} \approx 2.1 \times 10^{-18}$ J $\approx 13$ eV, so that $\delta \varepsilon = 1$ keV corresponds to $\varepsilon \approx 77$.

\begin{deluxetable*}{cccl}
\tablenum{1}
\tablecaption{
Common Simulation Parameters
\label{tab:parameters}
}
\tablewidth{0pt}
\tablehead{
\colhead{Parameter} & \colhead{Value} & \colhead{Unit} & \colhead{Name}
}
\startdata
$L_x$ & 200 & $d_p$ & Length in $\hat{x}$ \\
$\Delta x$ & 0.5 & $d_p$ & Cell size in $\hat{x}$ \\
$L_y$ & 64 & $d_p$ & Length in $\hat{y}$ \\
$\Delta y$ & 0.5 & $d_p$ & Cell size in $\hat{y}$ \\
$L_z$ & 64 & $d_p$ & Length in $\hat{z}$ \\
$\Delta z$ & 0.5 & $d_p$ & Cell size in $\hat{z}$ \\
$\tau$ & 60 & $\Omega_p^{-1}$ & Total time \\
$\Delta t$ & 0.01 & $\Omega_p^{-1}$ & Time step \\
$\delta t$ & 1 & $\Omega_p^{-1}$ & Output cadence \\
$N_{p0}$ & 50 & & Macro-particles per cell (initial) \\
$\beta_{e,i}$ & 1.0 & & Electron and proton plasma beta \\
\enddata
\end{deluxetable*}

This work presents results from two primary sets of seven 3-D simulation runs, each with a different upstream flow speed, $V_1$: one set used $V_1 = 4\ V_{A,1}$ and one used $V_1 = 2\ V_{A,1}$. Both primary sets comprised seven simulation runs with shock-normal angle, $\theta_{Bn}$, from $0^\circ$ to $90^\circ$ in increments of $15^\circ$. The purpose of these fourteen primary simulation runs was to examine the relative effects of $\theta_{Bn}$ and $V_1$ on proton energy spectra. A secondary set of twelve 3-D simulation runs with $V_1 = 4\ V_{A,1}$ provides additional insight into the dependence of the downstream spectral index on $\theta_{Bn}$.

Table \ref{tab:parameters} gives the parameters common to all simulation runs. The simulation normalizes lengths to the upstream proton inertial length, $d_p = V_{A,1}/\Omega_p$, and times to the upstream inverse proton gyrofrequency, $\Omega_p^{-1}$. The spatial dimensions are $L_x = 200\ d_p$, $L_y = 64\ d_p$, and $L_z = 64\ d_p$, with a grid-cell length of $0.5\ d_p$ in each dimension. Each run covers $\tau = 60\ \Omega_p^{-1}$ in time at a time step of $\Delta t = 0.01\ \Omega_p^{-1}$ and an output cadence of $\delta t = 100\ \Delta t = 1\ \Omega_p^{-1}$. The simulation injects protons at $x=0$ and reflects them at $x=L_x$, causing the resultant shock-normal vector to point in the $-\hat{x}$ direction. There are initially 50 macro-particles per cell; the combination of injection-reflection and periodic boundary conditions means that the total number of protons within the simulation volume only increases.

\subsection{Shock tracking and Mach number}
\label{sec:Shock tracking and Mach number}

\begin{figure}
    \centering
    \includegraphics[width=0.7\textwidth]{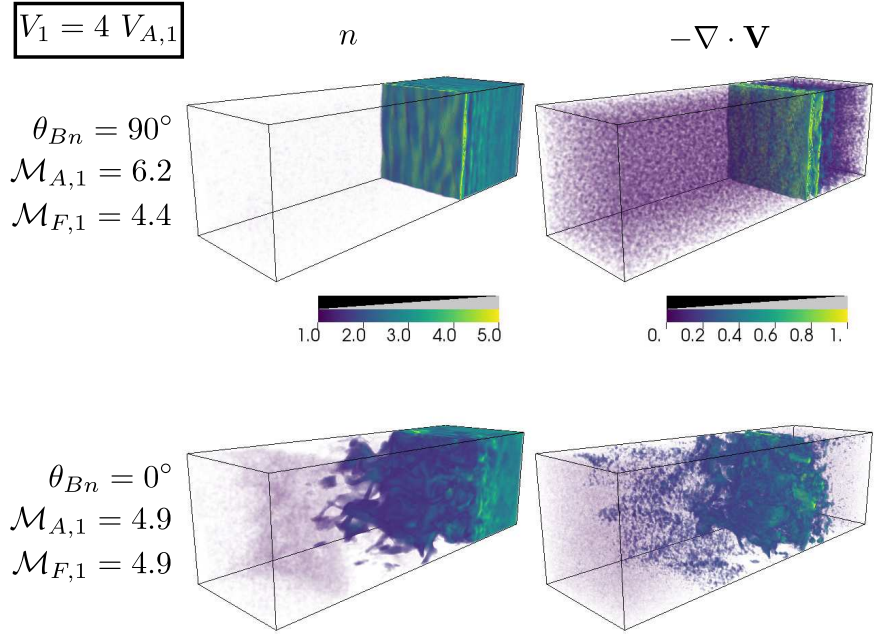}
    \caption{Density (left column) and velocity convergence (right column) during two simulation runs with $V_1 = 4\ V_{A,1}$. The top row shows images from the $\theta_{Bn} = 90^\circ$ simulation run and the bottom row shows images from the $\theta_{Bn} = 0^\circ$ simulation run. Both rows also list the upstream Alfv{\'e}n and fast magnetosonic Mach numbers, $\mathcal{M}_{A,1}$ and $\mathcal{M}_{F,1}$, as computed by the method in Section \ref{sec:Shock tracking and Mach number}.}
    \label{fig:den_cnvV-volumes-vx4}
\end{figure}

\begin{figure}
    \centering
    \includegraphics[width=0.7\textwidth]{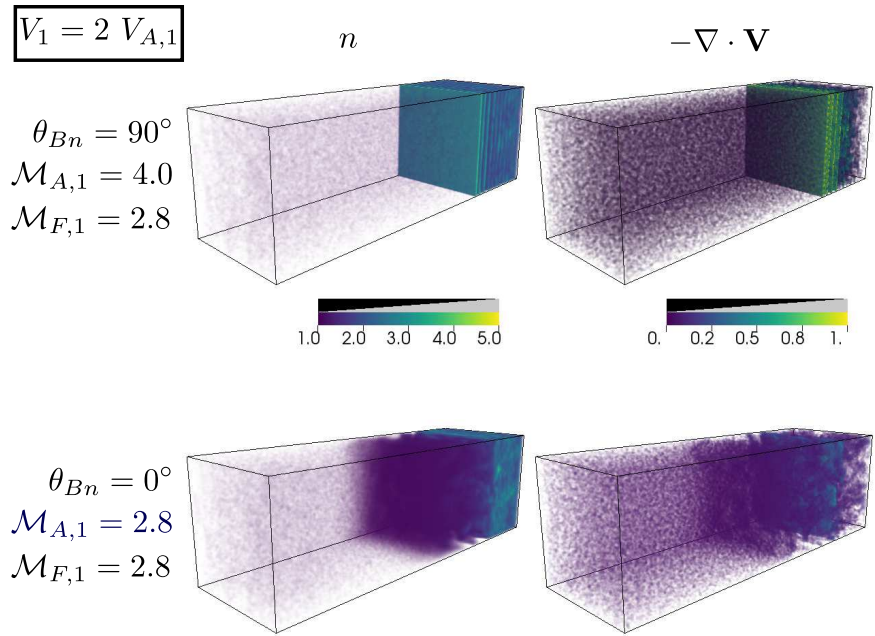}
    \caption{Same as Figure \ref{fig:den_cnvV-volumes-vx4}, except for runs with $V_1 = 2\ V_{A,1}$.}
    \label{fig:den_cnvV-volumes-vx2}
\end{figure}

Figure \ref{fig:den_cnvV-volumes-vx4} shows proton density, $n$, and velocity convergence, $-\nabla\cdot\mathbf{V}$ (i.e., negative velocity divergence), throughout the simulation volume  at $t = 30\ \Omega_p^{-1}$ during two simulation runs. The top row shows images from simulation run with $(\theta_{Bn}, V_1) = (90^\circ, 4\ V_{A,1})$ and the bottom row shows images from the simulation run with $(\theta_{Bn}, V_1) = (0^\circ, 4\ V_{A,1})$. The left column of images shows $n$ in normalized units from 1.0 to 5.0 and the right column shows $-\nabla\cdot\mathbf{V}$ in normalized units from 0.0 to 1.0. The advantage of showing $-\nabla\cdot\mathbf{V} \geq 0$ is that areas of relatively high positive convergence in the flow track shocks and strong compressions reasonably well \citep{Schwadron_etal-2015-ParticleAccelerationLow}. The shock front moves from back to front as time increases and is clear in images of both quantities. Figure \ref{fig:den_cnvV-volumes-vx2} shows the same quantities as Figure \ref{fig:den_cnvV-volumes-vx4}, at the same time step, for simulation runs with $(\theta_{Bn}, V_1) = (90^\circ, 2\ V_{A,1})$ and $(\theta_{Bn}, V_1) = (0^\circ, 2\ V_{A,1})$. Both shocks move more slowly than their $V_1 = 4\ V_{A,1}$ counterparts, as is to be expected, and both shock fronts exhibit less structure.

The upstream Alfv{\'e}n Mach number is defined as $\mathcal{M}_{A,1} \equiv V_s/V_{A,1}$, where $V_s$ is the shock speed in the upstream frame and $V_{A,1}$ is defined above. The upstream fast Mach number is similarly defined as $\mathcal{M}_{F,1} \equiv V_s/V_{F,1}$, where the upstream fast magnetosonic speed, $V_{F,1}$, depends on $V_{A,1}$, the upstream plasma thermal speed, $V_{th,1}$, and $\theta_{Bn}$:
\begin{displaymath}
    V_{F,1} = V_{A,1}\left\{\frac{1}{2}\left[1 + \beta_1 + \sqrt{\left(1 + \beta_1\right)^2 - 4\beta_1\cos^2\theta_{Bn}}\right]\right\}^{1/2}
\end{displaymath}
The quantity $\beta_1 = \left(V_{th,1}/V_{A,1}\right)^2$ here is equivalent to the upstream ratio of thermal to magnetic pressures. The simulation outputs $V_{th,1}^2(x,y,z,t)$ as a measure of total particle energy, and since the simulation normalizes all speeds to $V_{A,1}$, it automatically provides $\beta_1$ throughout the volume at each time step. This work used the average upstream value of $\beta_1$ at the first time step in order to compute $\mathcal{M}_{F,1}$. For reference, $\beta_1 \approx 1$ in these simulation runs, consistent with values in the solar wind near 1~au.

\begin{figure}
    \centering
    \includegraphics[width=0.5\textwidth]{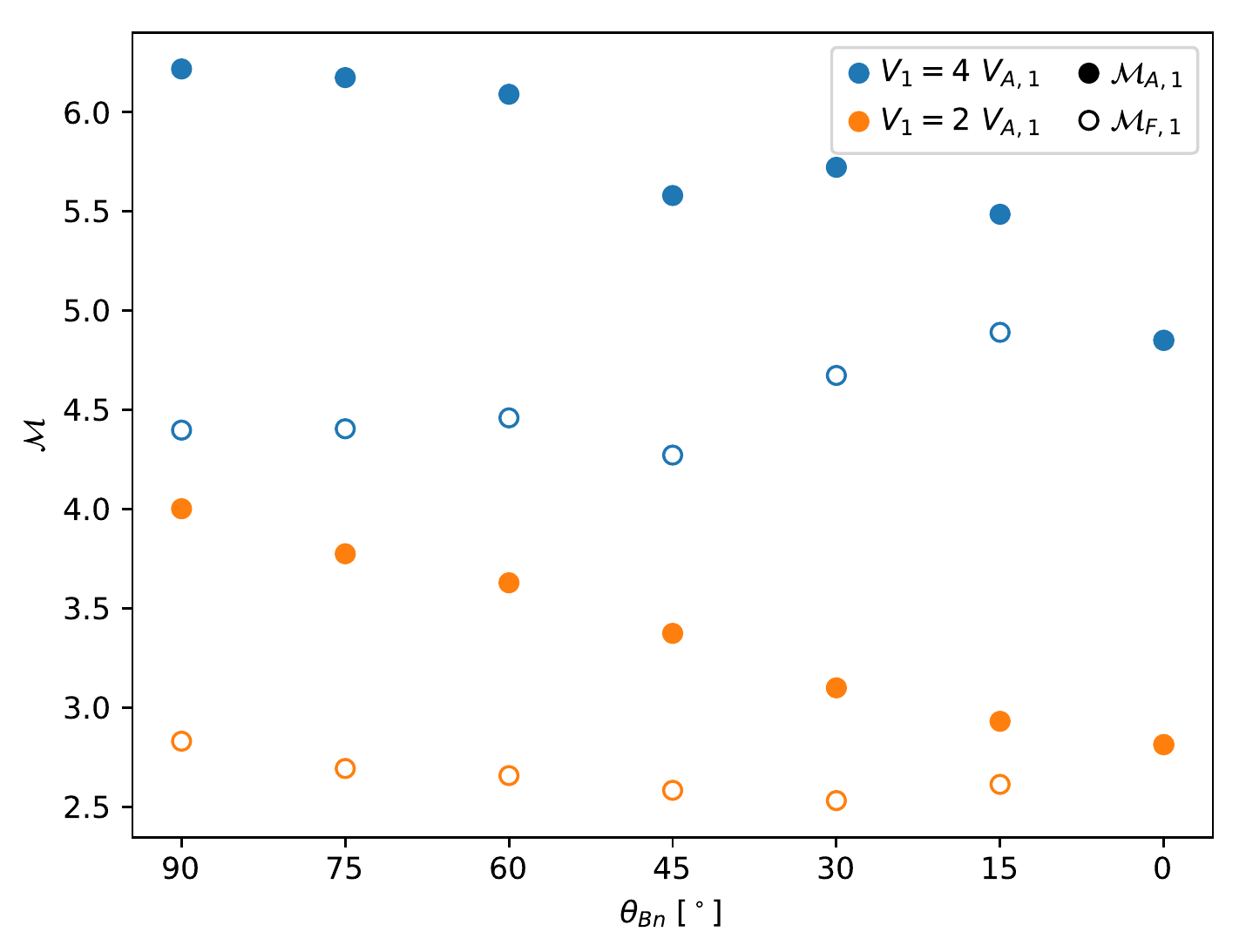}
    \caption{Upstream Alfv{\'e}n and fast Mach numbers, $\mathcal{M}_{A,1}$ and $\mathcal{M}_{F,1}$, of simulation runs as a function of shock-normal angle, $\theta_{Bn}$, grouped by upstream flow speed, $V_1$. Filled circles correspond to $\mathcal{M}_{A,1}$ and open circles correspond to $\mathcal{M}_{F,1}$. Note that the values of $\mathcal{M}_{A,1}$ and $\mathcal{M}_{F,1}$ are always greater at $4\ V_{A,1}$ than at $2\ V_{A,1}$ for each value of $\theta_{Bn}$.}
    \label{fig:mach_numbers-v1_groups}
\end{figure}

Figure \ref{fig:mach_numbers-v1_groups} shows $\mathcal{M}_{A,1}$ and $\mathcal{M}_{F,1}$ of the shock or compression in each simulation run, as estimated from tracking the $x$-axis position of maximum $-\nabla\cdot\mathbf{V}$. The values are shown as a function of shock-normal angle, $\theta_{Bn}$, and grouped by upstream flow speed, $V_1$. Our algorithm for estimating $\mathcal{M}_{A,1}$ from $-\nabla\cdot\mathbf{V}$ is as follows:
\begin{enumerate}
    \item Average $-\nabla\cdot\mathbf{V}$ over the $y$-$z$ plane.
    \item Find and store the index of the maximum value at each $\delta t$.
    \item Convert the index array to an array of physical position.
    \item Fit a line to the position array. The slope gives an estimate of the shock speed in the simulation reference frame.
    \item Transform the shock speed to the upstream reference frame.
\end{enumerate}
Since the index of maximum $-\nabla\cdot\mathbf{V}$ varied significantly during the first few $\Omega_p^{-1}$ before settling down, we computed a linear fit to the upstream shock speed after the first 10 $\Omega_p^{-1}$. Once we had a value for $\mathcal{M}_{A,1}$, we calculated $\mathcal{M}_{F,1}$ via the relation $\mathcal{M}_{F,1} = \mathcal{M}_{A,1}\left(V_{A,1}/V_{F,1}\right)$.

\section{Results} 
\label{sec:Results}

\subsection{Simulated data time series}
\label{sec:Simulated data time series}

\begin{figure}
    \centering
    \includegraphics[width=0.8\textwidth]{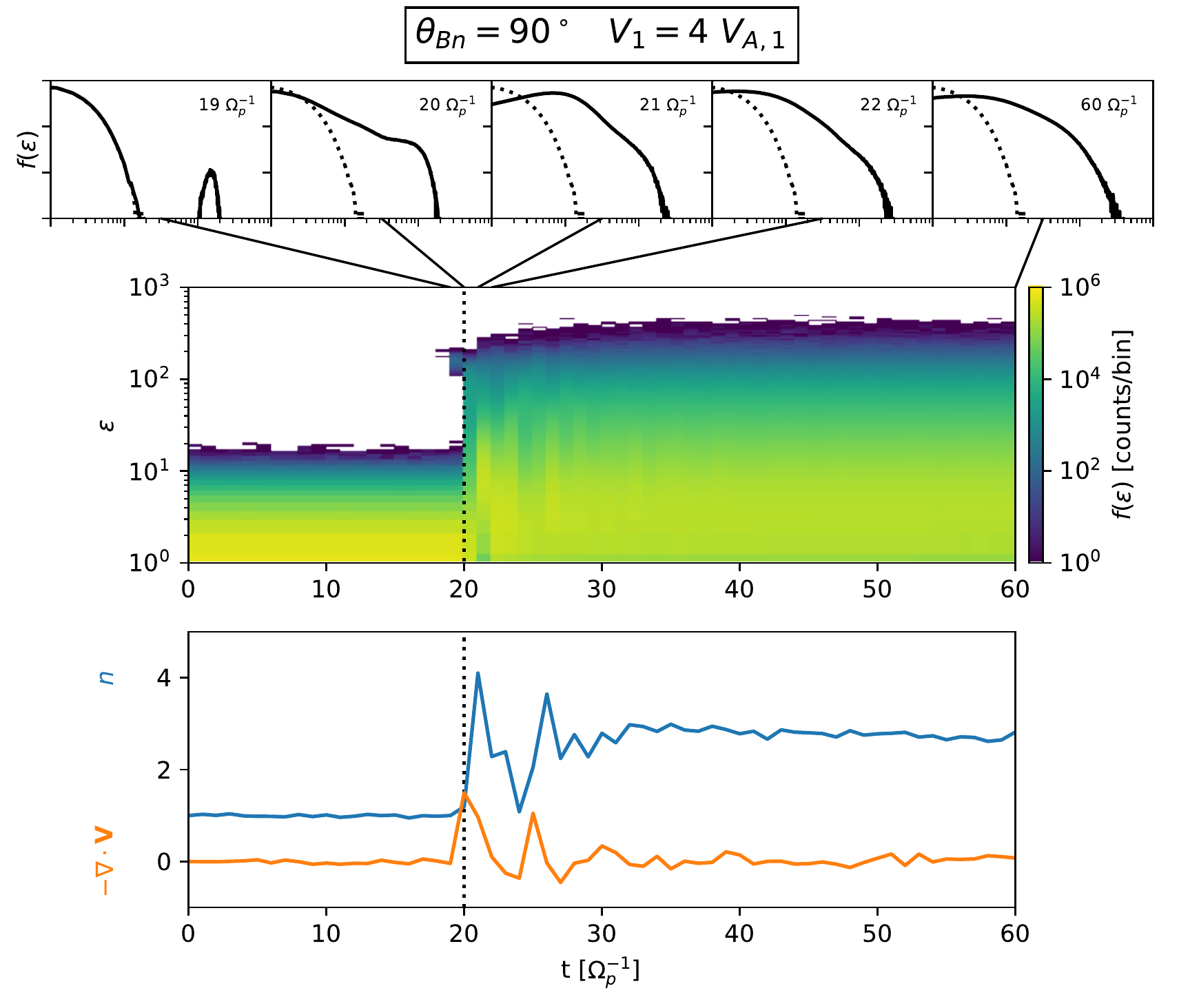}
    \caption{Time series from the simulation run with $\left(\theta_{Bn},V_1\right) = \left(90^\circ, 4\ V_{A,1}\right)$. \textit{Bottom:} The density, $n$, and velocity convergence, $-\nabla\cdot\mathbf{V}$, at $(x, y, z) = (3L_x/4, L_y/2, L_z/2)$. \textit{Middle:} The peculiar energy spectrogram, $f\left(\varepsilon, t\right)$, averaged over the $y$-$z$ plane and 1 $d_p$ at $x = 3L_x/4$. \textit{Top:} Snapshots of $f(\varepsilon)$ at specific times, as listed in each panel. The dotted line in each top-row panel shows the upstream distribution.}
    \label{fig:stackplot-timeseries-energy_slices-run000}
\end{figure}

Figure \ref{fig:stackplot-timeseries-energy_slices-run000} shows simulated time series of $n$, $-\nabla\cdot\mathbf{V}$, and the peculiar energy spectrum, $f\left(\varepsilon, t\right)$, for the $\theta_{Bn} = 90^\circ$ run. The $n$ and $-\nabla\cdot\mathbf{V}$ measurements in the bottom panel come from the point $(x, y, z) = (3L_x/4, L_y/2, L_z/2)$. The middle panel shows $f\left(\varepsilon\right)$ at each output step with color representing proton counts per bin. Each spectrum is averaged over a slab volume spanning the entire $y$-$z$ plane and one proton inertial length in the $x$ direction at $x = 3L_x/4$. This location captures downstream dynamics of even the slowest shocks while remaining far enough from the reflecting wall to avoid boundary effects. In both panels, a vertical dotted line marks the maximum value of $-\nabla\cdot\mathbf{V}$, which we take to be the time at which the shock passed this location.

The top row of panels shows $f\left(\varepsilon\right)$ at specific time steps chosen to illustrate the spectral evolution. The dotted line in each top-row panel represents the upstream thermal spectrum, for reference. The horizontal axis of each top-row panel spans $10^0 < \varepsilon < 10^3$, to match the vertical scale of the middle panel, and the common vertical axis spans $10^0$~counts~per~bin $< f\left(\varepsilon\right)$ $< 10^6$~counts~per~bin, to match the color scale of the middle panel. We have suppressed most axis annotations in the top row to reduce visual clutter.

The density in this perpendicular shock experiences a sharp increase to a factor of 4 at $20\ \Omega_p^{-1}$, coincident with the main compression. There is a smaller increase in $n$ from $19\ \Omega_p^{-1}$ to $20\ \Omega_p^{-1}$ which we take to be the shock ramp. Assuming the shock moves over this stationary fiducial observer at a speed of $2\ V_{A,1}$, the ramp is approximately $2\ d_p$ thick. Following the initial compression, there is a turbulent region of compressions and rarefactions before $n$ settles at an average downstream value of approximately 3 times greater than the upstream density. Although the compression ratio, $n_2/n_1 \approx 3$, is smaller than the value of 4 cited for canonically strong shocks, the value of $\mathcal{M}_{F,1} \approx 4.4$ shown in Figure \ref{fig:mach_numbers-v1_groups} makes this a relatively strong shock by IP standards.

Coincident with the passage of the $\theta_{Bn} = 90^\circ$ shock is an increase in protons with $\varepsilon \approx 100$, evident as an isolated patch of color to the left of the dotted line in the middle panel and as an isolated bump in the top-row spectrum at $19\ \Omega_p^{-1}$. As the shock passes, the initially Maxwellian distribution develops a wing that extends out to a few times $10\ \varepsilon$, similar to a $\kappa$ distribution. The suprathermal wing merges with the previously isolated burst at $\varepsilon \approx 100$, which has grown in amplitude. By $21\ \Omega_p^{-1}$, the spectral peak has shifted to $\varepsilon \approx 10$ and the $\varepsilon \sim 100$ population has begun to shrink. Up to $t \approx 30\ \Omega_p^{-1}$, the maximum energy increases until it reaches a roughly constant value of $\varepsilon \approx 300$, or $\delta \varepsilon \approx 3.9$ keV for typical 1-au values. During that period, the thermal component temporarily recovers before decreasing again, albeit to a lesser degree than just behind the shock. The spectrum from $40\ \Omega_p^{-1}$ through the end of the run is more or less stable, and its asymptotic form is a relatively smooth broken power law above $\varepsilon \sim 10$.

\begin{figure}
    \centering
    \includegraphics[width=0.8\textwidth]{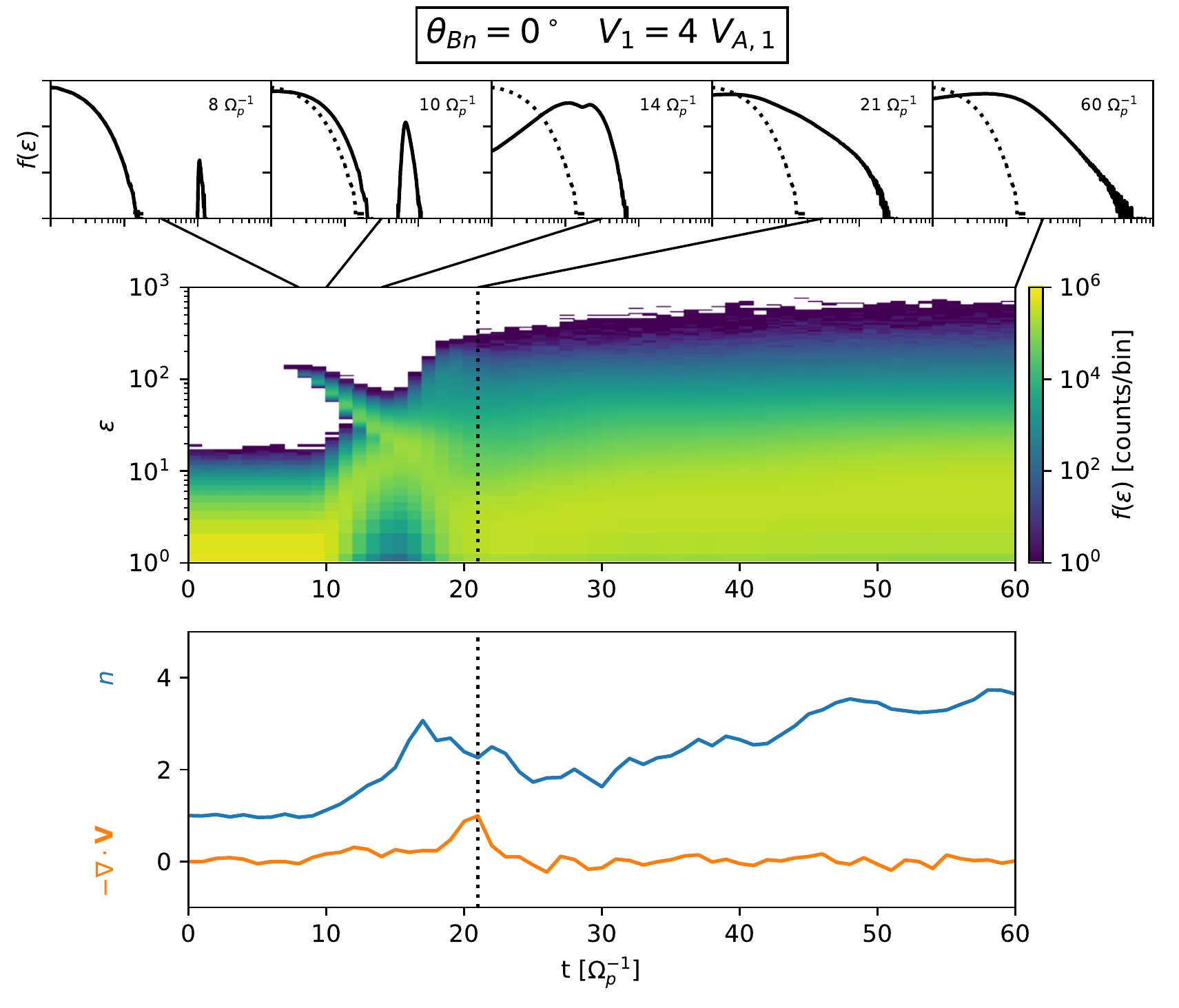}
    \caption{Time series from the simulation run with $\left(\theta_{Bn},V_1\right) = \left(0^\circ, 4\ V_{A,1}\right)$. The layout is the same as in Figure \ref{fig:stackplot-timeseries-energy_slices-run000}, except that the time step differs in some top-row panels.}
    \label{fig:stackplot-timeseries-energy_slices-run006}
\end{figure}

Figure \ref{fig:stackplot-timeseries-energy_slices-run006} shows simulated observations of $n$, $-\nabla\cdot\mathbf{V}$, and $f\left(\varepsilon, t\right)$, for the $\theta_{Bn} = 0^\circ$ run. The panel layout is identical to that of Figure \ref{fig:stackplot-timeseries-energy_slices-run000} but the times in the top panels differ in order to illustrate the distinct spectral evolution. The density in this parallel shock builds more gradually to a factor of 3 times the upstream value, then drops to just above 2 before the peak in $-\nabla\cdot\mathbf{V}$ passes. After the nominal shock passage, $n$ and $-\nabla\cdot\mathbf{V}$ continue to decrease until $30\ \Omega_p^{-1}$, at which point $n$ increases again. The downstream density increase is more gradual than the initial upstream increase and levels off between 3.5 and 4 near the end of the run. Based on the values of $n_2/n_1$ in this figure, and the value of $\mathcal{M}_{F,1} \approx 5.0$ in Figure \ref{fig:mach_numbers-v1_groups}, this also represents a strong IP shock.

The peculiar energy spectrum in the $\theta_{Bn} = 0^\circ$ run is markedly different in comparison to the $\theta_{Bn} = 90^\circ$ run. Like in the $\theta_{Bn} = 90^\circ$ run, there is an initial burst of protons with $\varepsilon \approx 100$ ahead of the shock. However, this suprathermal population appears much farther upstream than in the $\theta_{Bn} = 90^\circ$ run, since it is allowed to travel upstream along the magnetic field after acceleration and reflection at the shock. The initially narrow burst widens and its central energy decreases closer to the shock, forming a dispersive feature in $f\left(\varepsilon, t\right)$. The $f\left(\varepsilon\right)$ panels at $8\ \Omega_p^{-1}$ and $10\ \Omega_p^{-1}$ show this increase in width, as well as an increase in the bump-distribution amplitude, as the central energy decreases from $\varepsilon \approx 100$ to 70. The amplitude increase implies that the fraction of protons which the shock accelerates is inversely proportional to the energy they gain.

There is a second spectral feature that develops ahead of the shock. Between $10\ \Omega_p^{-1}$ and $15\ \Omega_p^{-1}$, the upstream thermal population decreases significantly as the peak energy increases to $\varepsilon \approx 10$. This, again, shares some similarities to the $\theta_{Bn} = 90^\circ$ run, but there are notable differences. One similarity is that the density peak arrives around the time of the shift in peak energy, implying that both the $\theta_{Bn} = 90^\circ$ and $\theta_{Bn} = 0^\circ$ shocks can accelerate the thermal population to modestly suprathermal values ($10 < \varepsilon < 100$). However, this bulk acceleration appears \emph{upstream} of the nominal shock in the $\theta_{Bn} = 0^\circ$ run whereas it occurs just behind the shock in the $\theta_{Bn} = 90^\circ$ run. The upshot for the $\theta_{Bn} = 0^\circ$ run is a significant dropout, then recovery, of the thermal population between $10\ \Omega_p^{-1}$ and $20\ \Omega_p^{-1}$, ahead of the shock.

Downstream of the $\theta_{Bn} = 0^\circ$ shock, the maximum energy gradually increases as in the $\theta_{Bn} = 90^\circ$ run, though it reaches $\varepsilon \approx 500$ (two-thirds higher) by $60\ \Omega_p^{-1}$. The $\theta_{Bn} = 0^\circ$ spectrum at $21\ \Omega_p^{-1}$ is very similar to the $\theta_{Bn} = 90^\circ$ spectrum at $22\ \Omega_p^{-1}$ but it evolves into a single power law by $60\ \Omega_p^{-1}$. Section \ref{sec:Final downstream energy spectra} compares and discusses final downstream spectra in greater detail.


\begin{figure}
    \centering
    \includegraphics[height=8in]{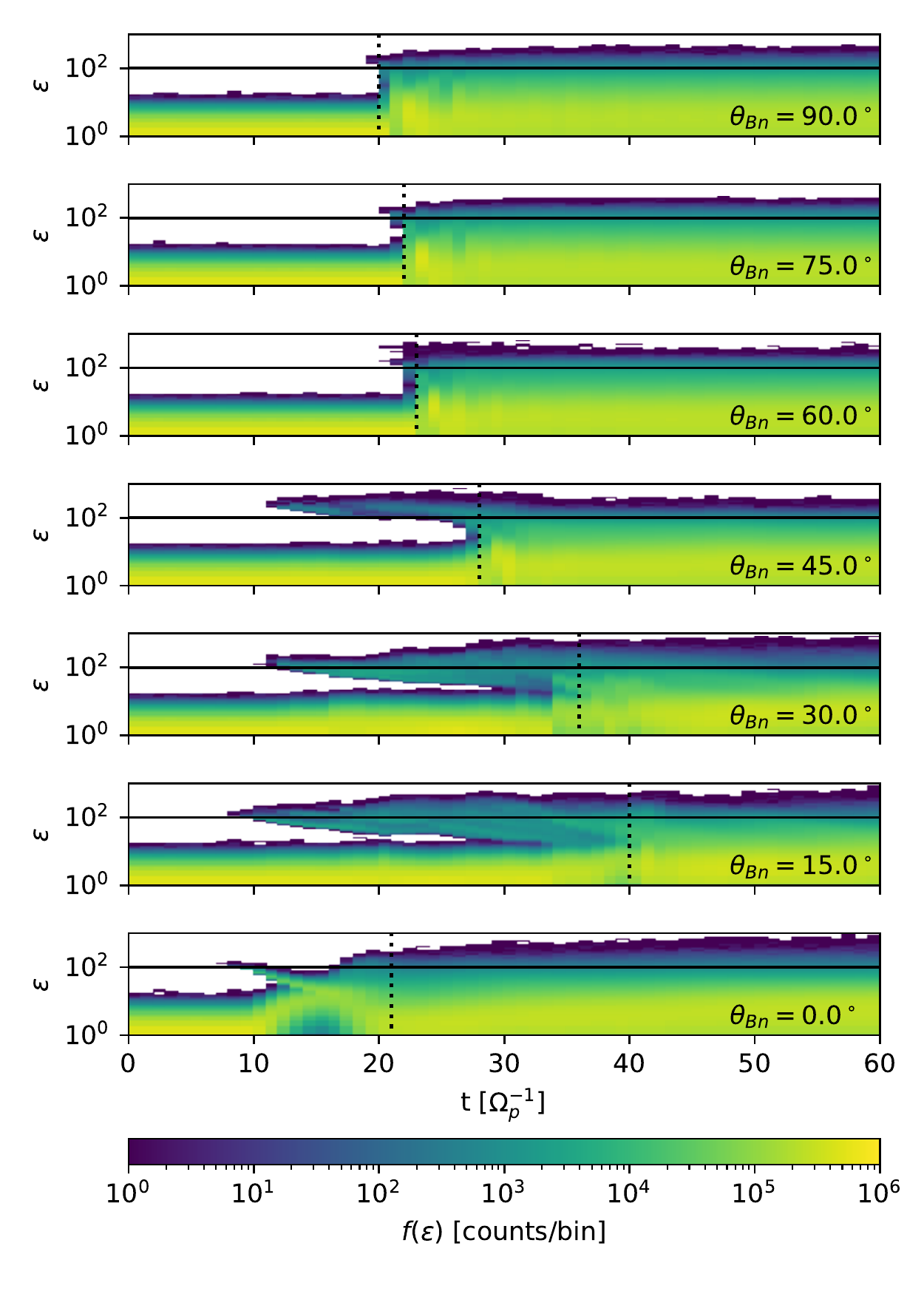}
    \caption{Proton peculiar energy spectrograms during all runs with $V_1 = 4\ V_{A,1}$. Time is shown in units of proton gyro cycles. Each panel lists the value of $\theta_{Bn}$ for that run. Vertical dotted lines mark $\left(-\nabla\cdot\mathbf{V}\right)_{max}$ as in Figures \ref{fig:stackplot-timeseries-energy_slices-run000} and \ref{fig:stackplot-timeseries-energy_slices-run006}.}
    \label{fig:epdistrh_2D-timeseries-vx4}
\end{figure}

Figure \ref{fig:epdistrh_2D-timeseries-vx4} shows $f\left(\varepsilon, t\right)$ during all seven runs with upstream speed $V_1 = 4\ V_{A,1}$. A solid horizontal line denotes $\varepsilon = 100$ and each panel lists the value of $\theta_{Bn}$ for that run in the lower right corner. Otherwise, the format is identical to that of the middle rows Figures \ref{fig:stackplot-timeseries-energy_slices-run000} and \ref{fig:stackplot-timeseries-energy_slices-run006}, which this figure repeats to facilitate comparison. This figure illustrates how the dynamic nature of $f\left(\varepsilon, t\right)$, averaged over a thin slab at $x = 3L_x/4$, varies with $\theta_{Bn}$.

The three runs with $\theta_{Bn} \geq 60^\circ$ are very similar. The major difference is that the drastic change in spectral shape associated with the shock arrives at $2\ \Omega_p^{-1}$ later for $\theta_{Bn} = 75^\circ$, and at $3\ \Omega_p^{-1}$ later for $\theta_{Bn} = 60^\circ$, compared to $\theta_{Bn} = 90^\circ$. Behind all three shocks, the maximum $\varepsilon$ increases steadily to a few hundred, though there is a very low amplitude group at high energies around the time of shock passage. This feature may be related to the horn-like spectral feature at smaller $\theta_{Bn}$, to be discussed below.

At $\theta_{Bn} = 45^\circ$, the behavior changes. The three runs with $45^\circ \geq \theta_{Bn} \geq 15^\circ$ include a horn-like shape that appears at $\varepsilon \sim 100$, in advance of the shock. The shock signature itself, in the form of an abrupt increase at all energies above the thermal population, is relatively clear around $28\ \Omega_p^{-1}$ for $\theta_{Bn} = 45^\circ$. The transition is more subtle for $\theta_{Bn} = 30^\circ$, with the spectrum filling in higher energies after the horn and thermal populations merge around $34\ \Omega_p^{-1}$. In the run with $\theta_{Bn} = 15^\circ$, the horn-like structure again merges with the thermal population around $34\ \Omega_p^{-1}$ but the spectrum between $\varepsilon \sim 10$ and $\varepsilon \sim 100$ fills in more gradually. 

The dispersive spike that appears at $\varepsilon \sim 100$ for $\theta_{Bn} = 0^\circ$ arrives around the same time as the horn-like structure in the run with $\theta_{Bn} = 15^\circ$. They and their counterparts in the $\theta_{Bn} = 30^\circ$ and $\theta_{Bn} = 45^\circ$ runs appear to both be the spectral signature of energetic protons returning back upstream from the shock. In the cases of $\theta_{Bn} = 15^\circ$, $30^\circ$, and $45^\circ$, the horn-like populations contain substructures in the form of quasi-periodic striations of decreasing $\varepsilon$. These substructures, which are most notable in the run with $\theta_{Bn} = 15^\circ$, suggest that the shocks with $\theta_{Bn} \leq 45^\circ$ repeatedly accelerate a fraction of the protons upstream as they propagate, and that those reflected beams appear as periodic dispersive structures when the fiducial observer is magnetically connected to the acceleration region. The higher number of macro-particles per cell in the simulation runs presented here, compared to previous 3-D hybrid simulations of proton acceleration at shocks, allows us to resolve this fine spectral structure.

In the special case of $\theta_{Bn} = 0^\circ$, reflected protons traveling exactly anti-parallel to $\hat{x}$ create the dispersive feature discussed above for Figure \ref{fig:stackplot-timeseries-energy_slices-run006}. The dynamic shape of the spectrum in the run with $\theta_{Bn} = 0^\circ$ is distinct from all the other six runs shown here, but additional runs at $\theta_{Bn} = 10^\circ$, $5^\circ$, and $1^\circ$ (not shown) suggest a continuous transition in which the thicker horn-like shape transitions to a narrow dispersive feature when $\theta_{Bn} \sim 1^\circ$.

Most aspects of $f\left(\varepsilon, t\right)$ do not significantly differ from $V_1 = 4\ V_{A,1}$ to $V_1 = 2\ V_{A,1}$; it will suffice to describe the relevant differences without showing additional spectra. For $90^\circ \geq \theta_{Bn} \geq 60^\circ$, the nearly discontinuous expansion out to higher energies occurs later, consistent with the later arrival of a slower shock. One notable difference is that the run with $(\theta_{Bn}, V_1) = (60^\circ, 2\ V_{A,1})$ contains an isolated increase in counts with $\varepsilon \sim 100$ that starts at $20\ \Omega_p^{-1}$, similar to the time at which the shock arrives and the spectrum expands in the run with $(\theta_{Bn}, V_1) = (60^\circ, 4\ V_{A,1})$. The isolated peak has relatively low amplitude compared to the thermal population and spans only a few energy bins, but it persists for roughly 5 $\delta t$. Similarly, the dynamic spectra for $\theta_{Bn} \leq 45^\circ$ do not significantly differ in overall form with decreasing $V_1$, except that their highest energies are proportionally lower. The most notable difference is that the horn-like feature in runs with $45^\circ \geq \theta_{Bn} \geq 15^\circ$ persists for longer before merging with the shocked population. For example: Whereas the horn-like structure appears at $11\ \Omega_p^{-1}$ and fully merges with the shocked population at $27\ \Omega_p^{-1}$ when $V_1 = 4\ V_{A,1}$, it extends from $15\ \Omega_p^{-1}$ to $38\ \Omega_p^{-1}$ when $V_1 = 2\ V_{A,1}$. This is due to the fact that the reflected ions have similar energies, and therefore speeds, in all runs, while shock speed decreases with $V_1$. One additional difference associated with decreasing $V_1$ is that the thermal population drop-out at $10\ \Omega_p^{-1} < t < 20\ \Omega_p^{-1}$ in the $\theta_{Bn} = 0^\circ$ run effectively disappears when $V_1 = 2\ V_{A,1}$.

\subsection{Final downstream energy spectra}
\label{sec:Final downstream energy spectra}

\begin{figure}
    \centering
    \includegraphics[width=1.0\textwidth]{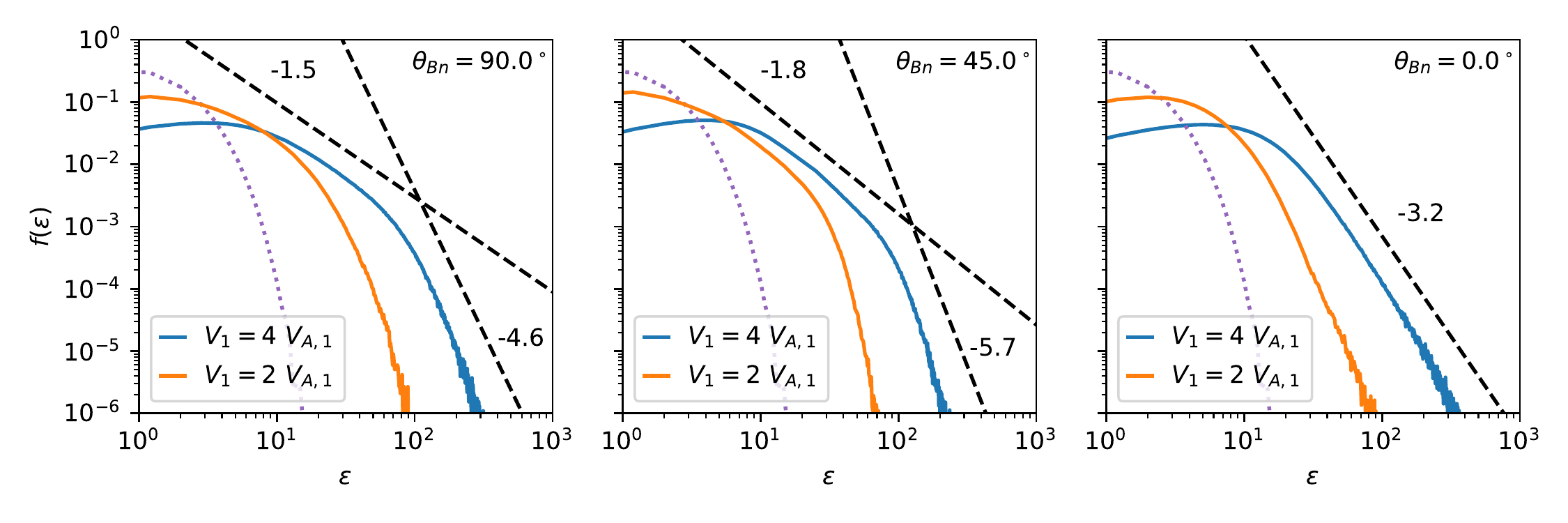}
    \caption{Normalized downstream energy spectra at the final time step in six runs. The spectra represent protons within a slab volume that spans the $y$-$z$ plane and one proton inertial length in $x$, located at $x = 3N_x/4$. Each panel shows the spectrum from the $V_1 = 4\ V_{A,1}$ run in blue and the spectrum from the $V_1 = 2\ V_{A,1}$ run in orange, as well as fiducial lines to illustrate power-law behavior in certain subdomains of $\varepsilon$. The dotted curve represents the initial thermal distribution. \textit{Left:} $\theta_{Bn} = 90^\circ$. \textit{Middle:} $\theta_{Bn} = 45^\circ$. \textit{Right:} $\theta_{Bn} = 0^\circ$. In all runs, the $2\ V_{A,1}$ spectra dominate up to $\varepsilon \approx 10$, at which point the the $4\ V_{A,1}$ spectra dominate.}
    \label{fig:epdistrh-downstream-combo}
\end{figure}

Figure \ref{fig:epdistrh-downstream-combo} shows downstream energy spectra at the last time step in example six runs. Each panel shows spectra from the $V_1 = 4\ V_{A,1}$ and $V_1 = 2\ V_{A,1}$ runs at a fixed value of $\theta_{Bn} \in \{0^\circ, 45^\circ, 90^\circ\}$, normalized to the total number of protons in the thin slab volume describe above. The dotted curve in each panel shows the initial thermal distribution for reference. The spectra with $\theta_{Bn} \in \{75^\circ, 60^\circ\}$ are very similar to those with $\theta_{Bn} = 90^\circ$ and spectra with $\theta_{Bn} \in \{30^\circ, 15^\circ\}$ are very similar to those with $\theta_{Bn} = 0^\circ$. Each panel also shows fiducial lines derived from power-law fits to regions of the $V_1 = 4\ V_{A,1}$ spectrum. These lines fit the data within a subdomain of $\varepsilon$ in a least-squares sense, and provide a reference spectrum elsewhere. In all runs, including those not shown, increasing upstream flow speed (i.e., upstream Mach number) significantly increases the fraction of protons with $\varepsilon \gtrsim 100$. This corresponds to protons that the shock has accelerated to roughly ten times the local thermal speed.

Both the $\theta_{Bn} = 90^\circ$ and $\theta_{Bn} = 45^\circ$ spectra with $V_1 = 4\ V_{A,1}$ have power-law forms, $f\left(\varepsilon\right) \propto \varepsilon^{-\gamma}$, for $\varepsilon \gtrsim 10$, with relatively smooth knees at $\varepsilon \approx 100$. The spectral index softens in both $\varepsilon$ subdomains as $\theta_{Bn}$ decreases from $\theta_{Bn} = 90^\circ$ to $\theta_{Bn} = 45^\circ$, and there is an accompanying decrease in maximum energy. Overlaying these two spectra shows that the $\theta_{Bn} = 45^\circ$ spectrum dominates the $\theta_{Bn} = 90^\circ$ in the vicinity of $\varepsilon = 10$, whereas the converse is true outside the subdomain $2 \lesssim \varepsilon \lesssim 20$. This may suggest that shocks with $\theta_{Bn} > 45^\circ$ more efficiently accelerate protons to $\varepsilon \sim 100$, or it may be the by-product of $\varepsilon \sim 100$ protons reflecting back upstream from the $\theta_{Bn} = 45^\circ$ shock and thus being absent from the downstream spectrum. 

The run with $\theta_{Bn} = 0^\circ$ also has a power-law tail, but has no spectral break around $\varepsilon = 100$. The lack of a spectral break causes the spectrum to extend to slightly higher $\varepsilon$ than the other two example spectra, but the compensating effect of the softer slope keeps the maximum value of $\varepsilon$ comparable across runs. The disappearance of the spectral break below $\theta_{Bn} = 45^\circ$ suggests that the $\varepsilon \sim 100$ protons accelerated at the shock and reflected back upstream are able to return downstream to contribute to the observed spectrum. 

The three $V_1 = 2\ V_{A,1}$ spectra are fundamentally similar in form to their respective $V_1 = 4\ V_{A,1}$ counterparts. The major differences are that they reach lower values of maximum $\varepsilon$ and contain a relatively greater thermal component. Both differences suggest that decreasing the Mach number while keeping $\theta_{Bn}$ fixed reduces the acceleration efficiency but does not change the acceleration process. It is worth noting that the $\left(\theta_{Bn}, V_1\right) = \left(90^\circ, 2\ V_{A,1}\right)$ spectrum has an exponential shape, $f\left(\varepsilon\right) \propto e^{-\varepsilon}$, making it appear more like a broadened Maxwellian than a broken power law. There is also no evidence of an isolated high-energy component in $f\left(\varepsilon, t\right)$ (not shown), unlike for the $V_1 = 4\ V_{A,1}$ run shown in Figures \ref{fig:stackplot-timeseries-energy_slices-run000} and \ref{fig:epdistrh_2D-timeseries-vx4}. In this case, the shock has essentially heated the distribution rather than accelerating a fraction of protons.

\begin{figure}
    \centering
    \includegraphics[width=0.5\textheight]{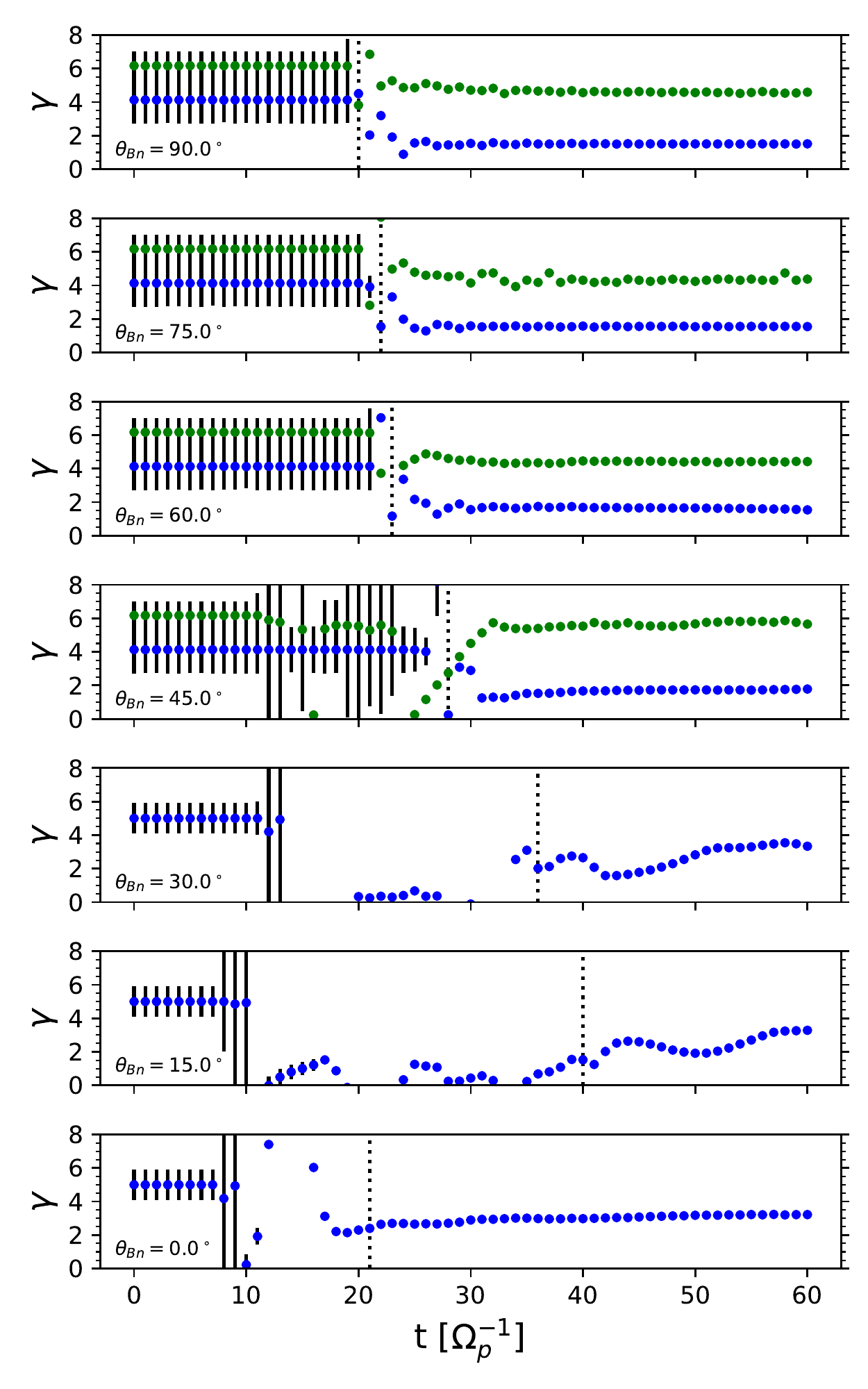}
    \caption{Spectral index fit as a function of time for each simulation run with $V_1 = 4\ V_{A,1}$. The arrangement is similar to that of Figure \ref{fig:epdistrh_2D-timeseries-vx4}: $\theta_{Bn}$ decreases from $90^\circ$ to $0^\circ$ from top to bottom and a dotted line indicates the time of shock passage. In the $\theta_{Bn} = 45^\circ$ to $90^\circ$ panels, green dots represent the high-energy spectral index while blue dots represent the low-energy spectral index.}
    \label{fig:epdistrh-xr50-vx4_original-slope-time}
\end{figure}

In order to study the evolution of $\gamma(t)$ leading up to the spectra in Figure \ref{fig:epdistrh-downstream-combo}, Figure \ref{fig:epdistrh-xr50-vx4_original-slope-time} shows fit values versus time for all seven runs with $V_1 = 4\ V_{A,1}$. The evolution of $\gamma(t)$ after shock passage is very similar in the $\theta_{Bn} = 90^\circ$, $75^\circ$, and $60^\circ$ simulation runs. After about five gyroperiods, the spectral index in the high-energy subdomain settles into its asymptotic value, $\gamma_{hi} \approx 4.5$, within one or two gyroperiods of the shock passage, while the spectral index in the low-energy subdomain varies about $\gamma_{lo} = 2$ for up to ten gyroperiods before settling to $\gamma_{lo} \approx 1.5$. 

The simulation run with $\theta_{Bn} = 45^\circ$ also resembles a broken power law, but the evolution of $\gamma_{hi}$ and $\gamma_{lo}$ differs slightly from the more perpendicular runs. The high index increases nearly linearly with time, beginning a few gyroperiods ahead of the shock and reaching an asymptotic value of $\gamma_{hi} \lesssim 6.0$ five gyroperiods after the shock passage. Just after the shock passage, $\gamma_{lo} \approx \gamma_{hi}$ for roughly one gyroperiod before quickly dropping to $\gamma_{lo} \gtrsim 1.0$, then slowly rising toward 2.0.

The $\theta_{Bn} = 45^\circ$ case represents the transition between predominantly perpendicular shocks and predominantly parallel shocks, which the evolution of $\gamma_{hi}$ illustrates particularly well. Whereas the effect of the shock on both $\gamma_{hi}$ and $\gamma_{lo}$ in the simulation runs with $\theta_{Bn} > 45^\circ$ is noticeable at most one gyroperiod ahead of the shock, consistent with Figure \ref{fig:epdistrh_2D-timeseries-vx4}, $\gamma_{hi}$ values begin to deviate from their pre-fit default values nearly twenty gyroperiods before the shock arrival. We must emphasize that the large error bars on these values indicate that they do not accurately represent the spectrum, which is far from a power law in the high $\varepsilon$ subdomain. However, their deviation from the default values corresponds to the emergence of the horn-like feature in Figure \ref{fig:epdistrh_2D-timeseries-vx4}.

Energy spectra in simulation runs with $\theta_{Bn} \leq 30^\circ$ are better fit by a single power law than a broken power law, so the corresponding panels show only a singly valued $\gamma(t)$. As indicated above, the spectral index begins to evolve far upstream of the shock as the horn-like feature develops. Again, these upstream values do not necessarily correspond to a formal power law but they do contain information about the spectrum in the $50 \lesssim \varepsilon \lesssim 100$ subdomain. For example: There are no points shown between $t = 14\ \Omega_p^{-1}$ and $t = 19\ \Omega_p^{-1}$ for the simulation run with $\theta_{Bn} = 30^\circ$ because $\gamma < 0$ there, corresponding to a positive slope at energies just below the peak in the horn-like feature. At $t = 20\ \Omega_p^{-1}$, $\gamma \approx 0$, corresponding to the development of a flat top as the horn-like feature spreads to lower energies. The simulation run with $\theta_{Bn} = 15^\circ$ undergoes a similar evolution, and $\gamma(t)$ varies between 2 and 3 before approaching a final value slightly above 3.

The evolution of $\gamma(t)$ for $\theta_{Bn} = 0^\circ$ just prior to the shock is qualitatively similar to the $\theta_{Bn} = 15^\circ$ and $\theta_{Bn} = 30^\circ$ cases, due to the narrow spectral feature shown in Figure \ref{fig:epdistrh_2D-timeseries-vx4}. However, it is far more stable after the shock passage than the other two, increasing steadily from 2.4 to 3.2. The behavior of $\gamma(t)$ for $\theta_{Bn} = 10^\circ$, $5^\circ$, and $1^\circ$ (not shown) is more similar to that for $\theta_{Bn} = 15^\circ$ and $30^\circ$, suggesting that this steady post-shock evolution turns on at $\theta_{Bn} = 0^\circ$.

\begin{figure}
    \centering
    \includegraphics[width=0.5\textwidth]{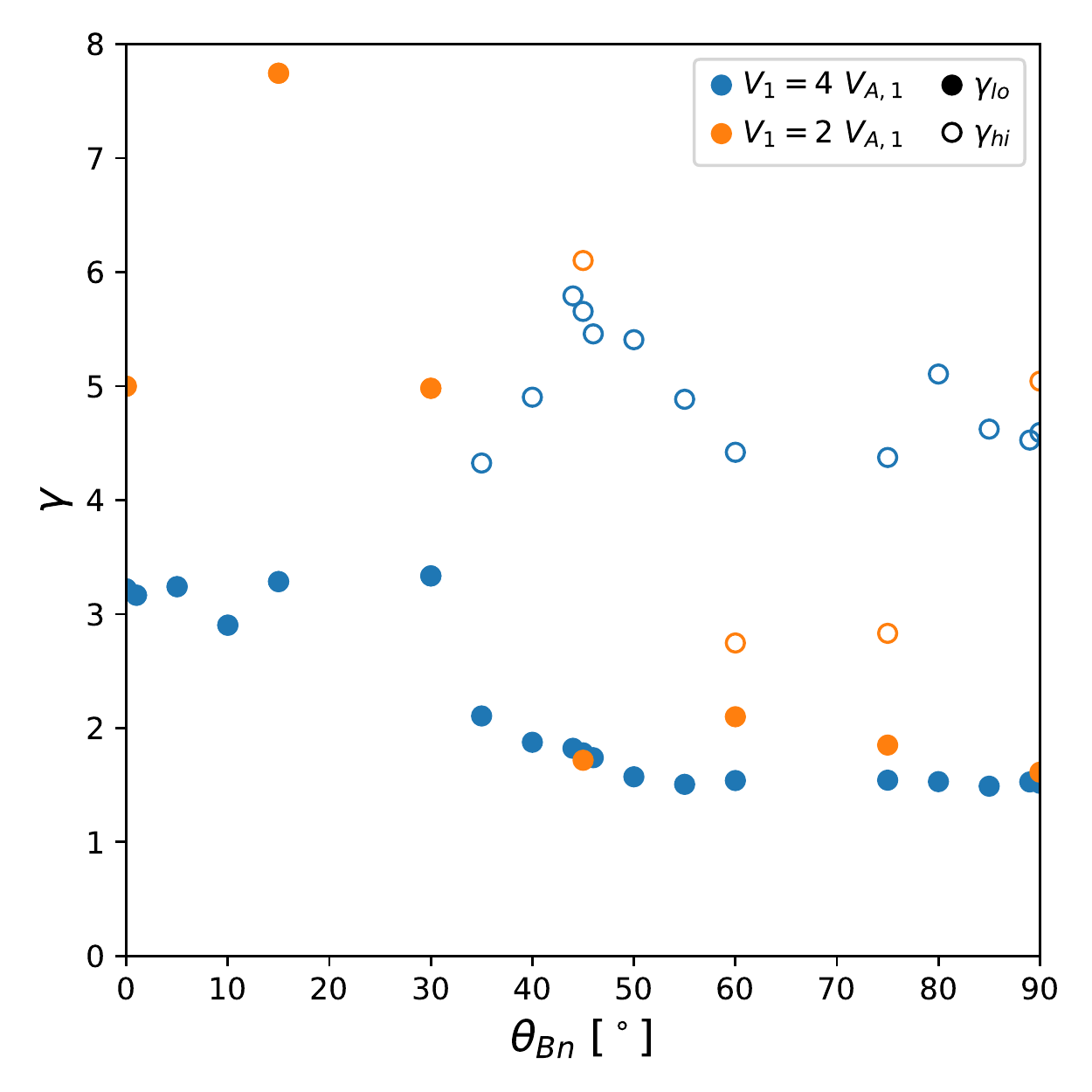}
    \caption{Spectral index fits at the final time step, as a function of $\theta_{Bn}$, for all simulation runs with $V_1 = 4\ V_{A,1}$ (filled circles) and $V_1 = 2\ V_{A,1}$ (open circles). The simulation runs with $0^\circ \leq \theta_{Bn} \leq 30^\circ$ have a single spectral index whereas the simulation runs with $35^\circ \leq \theta_{Bn} \leq 90^\circ$ have two spectral indices.}
    \label{fig:epdistrh-xr50-t60-vx2vx4-gamma-theta}
\end{figure}

Figure \ref{fig:epdistrh-xr50-t60-vx2vx4-gamma-theta} presents $\gamma(\theta_{Bn})$ at the final time step in all seven primary $V_1 = 4\ V_{A,1}$ runs (e.g., those shown in Figure \ref{fig:epdistrh-xr50-vx4_original-slope-time}), as well as the seven corresponding runs with $V_1 = 2\ V_{A,1}$ and a supplementary set of $V_1 = 4\ V_{A,1}$ runs at $\theta_{Bn} = 89^\circ$, $85^\circ$, $80^\circ$, $55^\circ$, $50^\circ$, $46^\circ$, $44^\circ$, $40^\circ$, $35^\circ$, $10^\circ$, $5^\circ$, and $1^\circ$. In the full set of $V_1 = 4\ V_{A,1}$ runs, $\gamma(0^\circ \leq \theta_{Bn} \leq 30^\circ) \approx 3$ and $\gamma_{lo}(60^\circ \leq \theta_{Bn} \leq 90^\circ) \approx 1.5$, consistent with previous figures, but the increased $\theta_{Bn}$ resolution indicates a sharp transition between $\theta_{Bn} = 30^\circ$ and $\theta_{Bn} = 35^\circ$, where $\gamma_{lo}$ drops to approximately 2 before further decreasing as $\theta_{Bn} \rightarrow 90^\circ$. The behavior of $\gamma_{hi}(\theta_{Bn} \geq 35^\circ)$ is more erratic than that of $\gamma_{lo}(\theta_{Bn} \geq 35^\circ)$, rapidly increasing to nearly 6 before returning to values between 4.5 and 5. 

The spectral indices at the end of $V_1 = 2\ V_{A,1}$ runs are much more highly variable than their $V_1 = 4\ V_{A,1}$ counterparts. The value of $\gamma_{lo}(\theta_{Bn} = 45^\circ)$ and $\gamma_{lo}(\theta_{Bn} = 90^\circ)$ are very similar at both speeds, and the values of $\gamma_{hi}(\theta_{Bn} = 45^\circ)$, $\gamma_{lo}(\theta_{Bn} = 60^\circ)$, $\gamma_{lo}(\theta_{Bn} = 75^\circ)$, and $\gamma_{hi}(\theta_{Bn} = 90^\circ)$ are within $\Delta \gamma = 1$ (i.e., an order of magnitude in energy), but the remaining values of $\gamma$ significantly differ between speeds. The most likely explanation for the high variability is that the $V_1 = 2\ V_{A,1}$ runs were still evolving at the end of the simulation time. This applies especially to the runs with $\theta_{Bn} = 15^\circ$ and $30^\circ$, whose shocks passed the simulated observer only $12\ \Omega_p^{-1}$ and $10\ \Omega_p^{-1}$, respectively, before the final time step. The indices for $V_1 = 2\ V_{A,1}$ transition between $\theta_{Bn} = 30^\circ$ and $\theta_{Bn} = 45^\circ$. Those with $\theta_{Bn} < 45^\circ$ are steeper than the corresponding $V_1 = 4\ V_{A,1}$ index while those with $\theta_{Bn} \geq 45^\circ$ are comparable.

\section{Discussion} 
\label{sec:Discussion}

One-dimensional simulations of quasi-parallel shocks by \citet{Kucharek_Scholer-1991-OriginDiffuseSuprathermal} and 3-D simulations of a parallel shock by \citet{Guo_Giacalone-2013-AccelerationThermalProtons} indicated that particles which are accelerated to high energies gained the initial energy boost at the shock front. 
Observational studies (e.g., by \citet{Kucharek_etal-2003-SourceAccelerationEnergetic}, \citet{Ebert_etal-2012-CorotatingInteractionRegion}, \citet{Filwett_etal-2017-SourcePopulationAcceleration}, and \citet{Lario_etal-2003-ACEObservationsEnergetic}) similarly suggest that CME- and CIR-associated shocks and compressions can locally accelerate suprathermal ions. The time-dependent spectra shown in Figure \ref{fig:epdistrh_2D-timeseries-vx4} support this picture of local shock acceleration in one way or another at all values of $\theta_{Bn}$: For shocks with $\theta_{Bn} > 45^\circ$, an isolated spike at or just ahead of the shock indicates drift-accelerated protons surfing along the shock front; for shocks with $\theta_{Bn} \leq 45^\circ$, an extended upstream beam of energetic protons indicates prior acceleration and reflection at the shock front. In contrast, there are no cases in which the spectrum rises gradually to higher energies downstream of the shock.

\citet{Lario_etal-2019-EvolutionSuprathermalProton} analyzed \textit{ACE} and \textit{Wind} data during the passage of IP shocks at 1~au, with a focus on relating suprathermal particle measurements to shock and upstream solar-wind parameters. They showed that the spacecraft observed beams of suprathermal protons upstream of parallel shocks, whereas they only observed suprathermal intensity enhancements very close to perpendicular shocks. Dynamic $1/v$ spectra for a shock with small $\theta_{Bn}$ near the point of observation on day 2001/304 show a dispersive feature similar to those in Figure \ref{fig:epdistrh_2D-timeseries-vx4} for $\theta_{Bn} = 15^\circ$ and $30^\circ$. In contrast, a bump in $\sim$ 8 keV protons appears just ahead of a shock on day 2001/118 with $\theta_{Bn} \approx 90^\circ$ near the spacecraft. This bump is more difficult to observe, but it is nonetheless consistent with the brief bumps around $\varepsilon \sim 100$ in Figure \ref{fig:epdistrh_2D-timeseries-vx4} for $\theta_{Bn} > 45^\circ$. Our simulation results self-consistently show how such behavior is a natural occurrence of shock acceleration of thermal protons, but we are cautious to not assume a direct correspondence between our stationary-frame simulations and moving-frame observations like those in \citet{Lario_etal-2019-EvolutionSuprathermalProton}.

Observations of the quiet solar wind proton velocity distribution consistently show a suprathermal power-law tail with spectral index of -5 \citep{Fisk_Gloeckler-2006-CommonSpectrumAccelerated} and multiple theories have arisen to explain the common $f(v) \propto v^{-5}$ distribution, which is equivalent to an energy distribution of the form $f(E) \propto E^{-5/2}$ or a differential intensity of the form $dJ/dE = 2Ef(E)/m^2 \propto E^{-3/2}$. Thermodynamic arguments in \citet{Fisk_Gloeckler-2006-CommonSpectrumAccelerated} spawned subsequent papers which further developed the theory of a pump acceleration mechanism to account for the $f(v) \propto v^{-5}$ form in ions with up to 100s of MeV throughout the heliosphere \citep{Fisk_Gloeckler-2007-AccelerationCompositionSolar, Fisk_Gloeckler-2008-AccelerationSuprathermalTails, Fisk_Gloeckler-2012-ParticleAccelerationHeliosphere, Fisk_Gloeckler-2014-CaseCommonSpectrum}. However, \citet{Schwadron_etal-2010-SuperpositionStochasticProcesses} showed that $f(v) \propto v^{-5}$ distributions can naturally arise from the superposition of stochastic processes which result from averaging observed spectra over significant time intervals, without the need for a universal acceleration mechanism. \citet{Jokipii_Lee-2010-CompressionAccelerationAstrophysical} pointed out critical short-comings in the theory of \citet{Fisk_Gloeckler-2007-AccelerationCompositionSolar} and \citet{Fisk_Gloeckler-2008-AccelerationSuprathermalTails}, and suggested instead that the quiet-time suprathermal population comprises remnant ions from various energetic particle events.

An energy spectrum with the power-law dependence $f(E) \propto E^{-\gamma}$ corresponds to a differential intensity with the dependence $dJ/dE \propto E^{-(\gamma-1)}$. The spectra in runs with $\theta_{Bn} \leq 30^\circ$ and $V_1 = 4\ V_{A,1}$ would therefore produce observations of $dJ/dE \propto E^{-2.2}$, which is close to the upper bound of values observed by \citet{Dayeh_etal-2009-CompositionSpectralProperties} and \citet{Desai_etal-2010-OriginQuietTime}. Similar observational results of $dJ/dE \propto E^{-2.51}$ \citep{Mason_etal-2008-AbundancesEnergySpectra} and $dJ/dE \propto E^{-2}$ \citep{Mewaldt_etal-2001-LongTermFluences} suggest that the acceleration mechanism in our simulation runs with $\theta_{Bn} \leq 30^\circ$, which is likely a first-order Fermi mechanism, most closely mimics the process that produces suprathermal ions downstream of shocks and compressions in the solar wind.

\citet{Caprioli_Spitkovsky-2014-SimulationsIonAcceleration1} found that fast parallel shocks in their hybrid simulations accelerated protons most efficiently and that acceleration efficiency as a function of $\theta_{Bn}$ dropped sharply around $45^\circ$ (cf. their Figure 3). Although the Mach numbers in their simulation runs tended to be much higher than in ours, the downstream spectra in Figure \ref{fig:epdistrh-downstream-combo} suggest a similar trend at $\varepsilon \gtrsim 200$: The softer index of the $\theta_{Bn} = 0^\circ$ downstream spectrum above $\varepsilon \approx 100$ leads to larger normalized $f(\varepsilon)$ (i.e., relatively more protons) at $\varepsilon = 200$ than in the other two runs.

\begin{figure}
    \centering
    \includegraphics[width=0.7\textwidth]{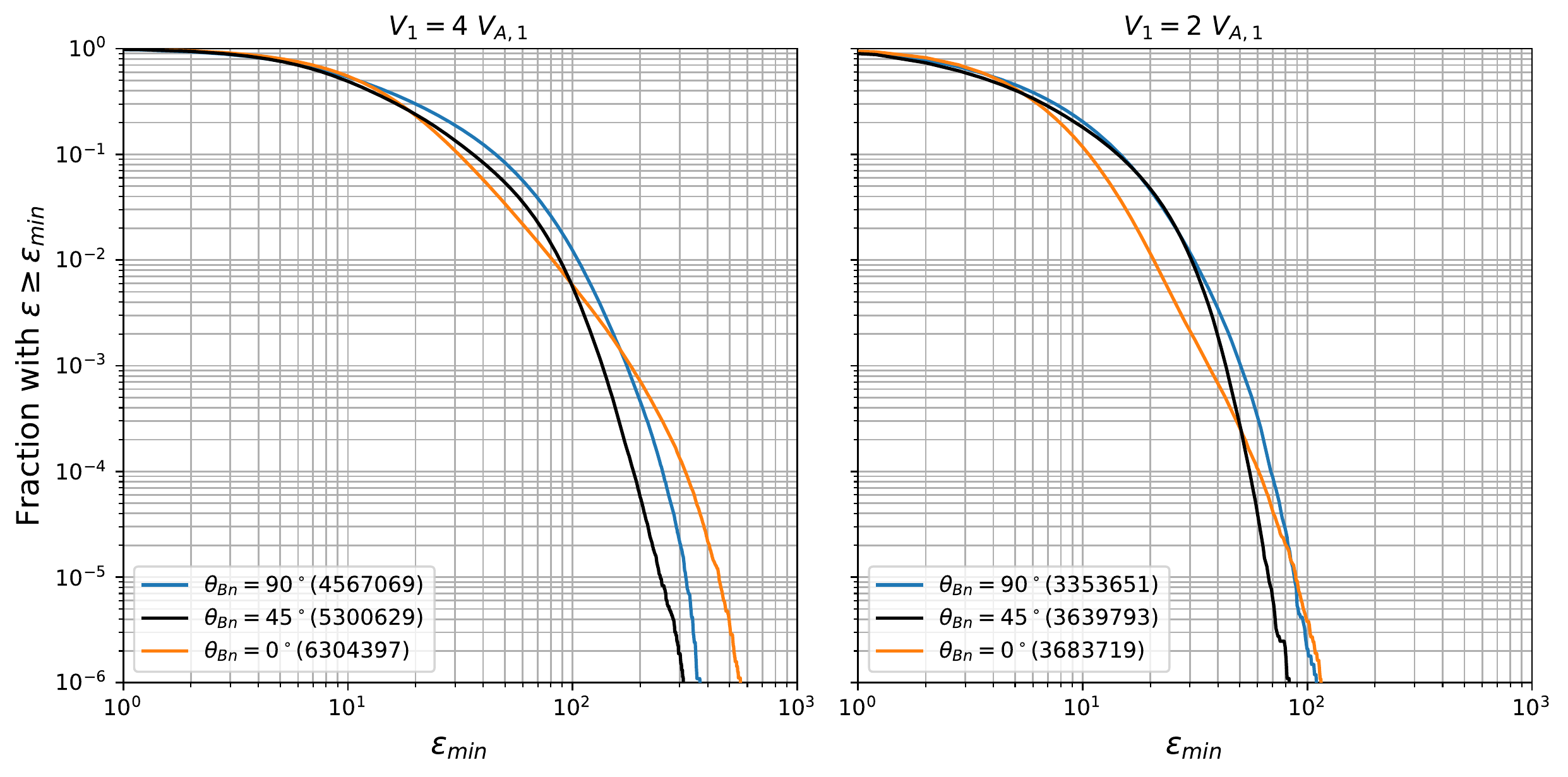}
    \caption{The fraction of downstream protons with energy equal to or greater than each energy bin in the simulation runs shown in Figure \ref{fig:epdistrh-downstream-combo}. Each spectrum is normalized to the number of protons (listed in the legend) that it represents.}
    \label{fig:epdistrh-vx4vx2_acceleration_efficiency}
\end{figure}

Figure \ref{fig:epdistrh-vx4vx2_acceleration_efficiency} shows the fraction of protons in each spectrum from Figure \ref{fig:epdistrh-downstream-combo} that have $\varepsilon \geq \varepsilon_{min}$, where $\varepsilon_{min}$ is the energy in each progressively higher energy bin. Each point on a particular curve thus gives the fraction of protons in the corresponding downstream spectrum with at least as much energy as $\varepsilon_{min}$. In the left panel ($V_1 = 4\ V_{A,1}$), the $\theta_{Bn} = 0^\circ$ curve lies below both other curves for $20 \lesssim \varepsilon_{min} \lesssim 100$, at which point it overtakes the $\theta_{Bn} = 45^\circ$ curve. It then overtakes the $\theta_{Bn} = 90^\circ$ curve at $\varepsilon_{min} \approx 150$ and remains above both out to the highest energies. 

In the right panel ($V_1 = 2\ V_{A,1}$), the $\theta_{Bn} = 0^\circ$ curve lies below both other curves for $6 \lesssim \varepsilon_{min} \lesssim 50$, at which point it overtakes the $\theta_{Bn} = 45^\circ$ curve. It follows the $\theta_{Bn} = 90^\circ$ curve out to $\varepsilon_{min} \approx 100$ but does not decisively overtake it. The shape of the $\theta_{Bn} = 0^\circ$ curve in both panels --- namely, the linear slope in the $\varepsilon_{min} \sim 10$ subdomain --- indicates that the parallel shock preferentially accelerates protons to relatively high energies. In contrast, the exponential shape of the other two runs indicates that they more evenly distribute protons among energies. A possible exception to this latter observation is the roughly linear dependence on $\varepsilon_{min}$ of the $(\theta_{Bn}, V_1) = (45^\circ, 4\ V_{A,1})$ curve at $\varepsilon_{min} > 100$. In this sense, the behavior of the $(\theta_{Bn}, V_1) = (45^\circ, 4\ V_{A,1})$ simulation run represents a transitionary case between the other two, consistent with results shown in previous figures.

The $V_1 = 4\ V_{A,1}$ panel in Figure \ref{fig:epdistrh-vx4vx2_acceleration_efficiency} illustrates an interesting dichotomy between perpendicular and parallel shocks with respect to proton acceleration. On one hand, the fact that the $(\theta_{Bn}, V_1) = (0^\circ, 4\ V_{A,1})$ curve dominates at $\varepsilon_{min} \geq 150$ indicates that the parallel shock preferentially accelerates protons to the highest energies observed in these simulation runs. On the other hand, the fact that the $(\theta_{Bn}, V_1) = (90^\circ, 4\ V_{A,1})$ curve dominates at $20 \leq \varepsilon_{min} \leq 100$ indicates that the perpendicular shock accelerates a greater fraction of protons to energies above the thermal population. While predominantly parallel shocks may be more efficient, predominantly perpendicular shocks seem to be more effective.

The former point is a well known consequence of DSA. To clarify the latter point, note that there are twice as many protons with $\varepsilon_{min} = 30$ downstream of the $(\theta_{Bn}, V_1) = (90^\circ, 4\ V_{A,1})$ shock than downstream of the $(\theta_{Bn}, V_1) = (0^\circ, 4\ V_{A,1})$ shock. In the $V_1 = 2\ V_{A,1}$ case, the ratio is five to one. Since $V_{A,1} = V_{th,1}$ in the simulation runs presented here, an energy of $\varepsilon = 30$ corresponds to speeds between 5 and 6 times $V_{th,1}$ in the frame of the solar wind. This means that these protons may have sufficient momentum to participate in DSA if they encounter a predominantly parallel shock or compression relatively soon afterward. Our results confirm that predominantly parallel shocks play a significant role in isolated proton acceleration events, but they also suggests that predominantly perpendicular shocks play an important role in conditioning a plasma volume for the effects of a subsequent shock, such as in CME-CME interactions.

As with Figure \ref{fig:epdistrh-downstream-combo}, the curves in Figure \ref{fig:epdistrh-vx4vx2_acceleration_efficiency} for $\theta_{Bn} \in \{75^\circ, 60^\circ\}$ are very similar to those for $\theta_{Bn} = 90^\circ$ and the curves for $\theta_{Bn} \in \{30^\circ, 15^\circ\}$ are very similar to those for $\theta_{Bn} = 0^\circ$. Our results therefore qualitatively agree with those of \citet{Caprioli_Spitkovsky-2014-SimulationsIonAcceleration1}. However, the spectra shown in Figure \ref{fig:epdistrh-downstream-combo}, and used to create Figure \ref{fig:epdistrh-vx4vx2_acceleration_efficiency}, represent only the protons downstream at the end of the run. They do not account for protons which remain near the shock, or which leave the local plasma due to a transport process. A thorough understanding of the fate of protons (and charged particles in general) energized at shocks should include those effects.

\textit{In situ} observations have shown that accelerated particles are often associated with shocks \citep{Giacalone-2012-EnergeticChargedParticles} but not all shocks accelerate particles \citep{Lario_etal-2003-ACEObservationsEnergetic}. \citet{Reames-2012-ParticleEnergySpectra} proposed that the absence of accelerated He ions in $\sim$85\% of 258 IP shocks observed by \textit{Wind} could be due to a low shock speed, a low shock compression ratio, or a small value of $\theta_{Bn}$. Our results clearly show that shock speed plays a significant role in accelerating thermal protons to suprathermal energies. In each pair of runs with a particular value of $\theta_{Bn}$, the spectrum downstream of the faster shock dominates the spectrum downstream of the slower shock for energies $\varepsilon \gtrsim 10$, and vice versa at $\varepsilon \lesssim 10$. That means that there are proportionally fewer thermal protons and more suprathermal protons downstream of the faster shock. The production of suprathermal protons depends more on shock speed than on $\theta_{Bn}$.

Finally, we note that injecting upstream magnetic-field turbulence increased the acceleration efficiency of thermal protons at perpendicular shocks in 2-D hybrid simulations by \citet{Giacalone-2005-EfficientAccelerationThermal} and at parallel shocks in 2-D hybrid/test-particle simulations by \citet{Guo_Giacalone-2012-ParticleAccelerationFlare}. Our simulation runs do not contain any upstream magnetic turbulence, but the downstream spectra at $60\ \Omega_p^{-1}$ in our runs with $\theta_{Bn} \geq 45^\circ$ and $V_1 = 4\ V_{A,1}$ are qualitatively similar to the downstream spectra at $80\ \Omega_p^{-1}$ in 3-D hybrid simulations by \citet{Giacalone_Ellison-2000-ThreeDimensionalNumerical} with $\theta_{Bn} = 80^\circ$ and $V_1 = 4\ V_{A,1}$, both with and without injected upstream magnetic-field turbulence. While future code developments may allow for injecting \textit{ad hoc} upstream turbulence, wave-particle interactions in the predominantly parallel simulation runs presented here generate turbulence that self-consistently influences the proton spectra.

\section{Conclusion} 
\label{sec:Conclusion}

This work presented hybrid numerical simulations of proton dynamics in moderate to strong collisionless IP shocks in the full range of shock-normal angles, $0^\circ \leq \theta_{Bn} \leq 90^\circ$, and at two values of the upstream flow speeds, $V_1 = 2\ V_{A,1}$ and $4\ V_{A,1}$. These combinations of $\theta_{Bn}$ and $V_1$ produced Alfv{\'e}n Mach numbers in the range $3.0 \lesssim \mathcal{M}_{A,1} \lesssim 6.0$, or fast magnetosonic Mach numbers in the range $2.5 \lesssim \mathcal{M}_{F,1} \lesssim 5.0$, typical of moderate to strong CME-driven shocks at 1 au.

The main conclusions of this paper are:
\begin{enumerate}
    \item The shock-normal angle, $\theta_{Bn}$, broadly organizes the spectral shape while the Mach number, $\mathcal{M}_{A,1}$ or $\mathcal{M}_{F,1}$, controls acceleration efficiency for a given run.
    \item All strong IP shocks accelerate some protons to hundreds of times the solar-wind thermal energy. Even moderate IP shocks can readily accelerate protons to a few tens of times the solar-wind thermal energy.
    \item Proton energy spectra downstream of strong IP shocks with $\theta_{Bn} \geq 35^\circ$ resemble broken power laws with break energy $\varepsilon \lesssim 100$. The lower spectral index is in the range $1.5 \leq \gamma_{lo} \leq 2.0$ and the higher spectral index is in the range $4.5 \leq \gamma_{hi} \leq 6.0$.
    \item Proton energy spectra downstream of strong IP shocks with $\theta_{Bn} \leq 30^\circ$ resemble single power laws. The spectral index is in the range $2.0 \leq \gamma \leq 4.0$ but is more highly variable over the course of a given simulation run than for shocks with $\theta_{Bn} \geq 35^\circ$.
    \item Spectral indices downstream of strong IP shocks at the end the simulation time tend to cluster around $\gamma \approx 3$ for $0^\circ \leq \theta_{Bn} \leq 30^\circ$. At $\theta_{Bn} = 35^\circ$, the lower spectral index drops to $\gamma_{lo} \approx 2$ and decreases nearly linearly with $\theta_{Bn}$ until settling at $\gamma_{lo} \approx 1.5$ for $55^\circ \leq \theta_{Bn} \leq 90^\circ$. The higher spectral index varies within $4 < \gamma_{hi} < 6$ for $\theta_{Bn} \geq 35^\circ$.
    \item Predominantly parallel shocks ($\theta_{Bn} \leq 30^\circ$) accelerate protons to higher energies than predominantly perpendicular shocks ($\theta_{Bn} \geq 60^\circ$), but predominantly perpendicular shocks produce more downstream suprathermal protons than predominantly parallel shocks.
\end{enumerate}

Although much of the variation in observations of the suprathermal solar wind is most apparent in composition or abundance effects \citep{Desai_etal-2006-HeavyIonElemental}, the results shown herein provide insight into the origin of the suprathermal seed population in the solar wind. This work also provides candidate seed spectra that can be used as initial conditions in particle acceleration and transport codes, such as the Energetic Particle Radiation Environment Model (EPREM) \citep{Schwadron_etal-2010-EarthMoonMars, Kozarev_etal-2013-GlobalNumericalModeling}, or which can be adapted to conditions closer to the Sun and compared to measurements from the Integrated Science Investigation of the Sun (IS$\Sun$IS) instrument suite onboard \textit{Parker Solar Probe} \citep{McComas_etal-2016-IntegratedScienceInvestigation,McComas_etal-2019-ProbingEnergeticParticle}.

\acknowledgments
NASA grant 80NSSC17K0009 to the University of New Hampshire (UNH) supported the work presented here. M. Young thanks M. Dayeh, R. Ebert, and R. Filwett for guidance regarding past and present research on suprathermal seed populations. Computations were performed on Trillian, a Cray XE6m-200 supercomputer at UNH supported by the NSF MRI program under grant PHY-1229408. This work used no public data.

%

\vspace{5mm}







\bibliography{myoung}
\bibliographystyle{aasjournal}


\end{document}